\documentclass[sigconf]{acmart}
\usepackage{subcaption}

\setcopyright{none} 
\acmConference[Anonymous Submission to ACM CCS 2021]{ACM Conference on Computer and Communications Security}{Due 6 May 2021}{Seoul, TBD}
\acmYear{2021}

\begin{document}
\title{(Un)clear and (In)conspicuous: The right to opt-out of sale \\under CCPA} 

\author{Sean O'Connor}
\affiliation{%
  \institution{Pomona College}
}
\email{swow2015@mymail.pomona.edu}

\author{Ryan Nurwono}
\affiliation{%
  \institution{UC Berkeley}
}
\email{ryanjnurwono@gmail.com}

\author{Aden Siebel}
\affiliation{%
  \institution{Pomona College}
}
\email{amsa2017@mymail.pomona.edu}

\author{Eleanor Birrell}
\affiliation{%
  \institution{Pomona College}
}
\email{eleanor.birrell@pomona.edu}

\begin{abstract}
The California Consumer Privacy Act (CCPA)---which began enforcement on July 1, 2020---grants California users the affirmative right to opt-out of the sale of their personal information. In this work, we perform a series of observational studies to understand how websites implement this right. We perform two manual analyses of the top 500 U.S. websites (one conducted in July 2020 and a second conducted in January 2021) and classify how each site implements this new requirement. We also perform an automated analysis of the Top 5000 U.S. websites. We find that the vast majority of sites that implement opt-out mechanisms do so with a Do Not Sell link rather than with a privacy banner, and that many of the linked opt-out controls exhibit features such as nudging and indirect mechanisms (e.g., fillable forms). We then perform a pair of user studies with 4357 unique users (recruited from Google Ads and Amazon Mechanical Turk) in which we observe how users interact with different opt-out mechanisms and evaluate how the implementation choices we observed---exclusive use of links, prevalent nudging, and indirect mechanisms---affect the rate at which users exercise their right to opt-out of sale. We find that these design elements significantly deter interactions with opt-out mechanisms---including reducing the opt-out rate for users who are uncomfortable with the sale of their information---and that they reduce users' awareness of their ability to opt-out. Our results demonstrate the importance of regulations that provide clear implementation requirements in order empower users to exercise their privacy rights.
\end{abstract}



\maketitle

\section{Introduction}

In recent years, growing recognition of the potential for exploitation of personal data and of the shortcomings of prior privacy regimes has led to the passage of new online privacy regulations, notably the EU's General Data Protection Regulation (GDPR)---which went into effect on May 25, 2018---and the California Consumer Privacy Act (CCPA)---which began enforcement on July 1, 2020. Several pieces of work have looked at the effect of GDPR on privacy policies and data practices~\cite{utz2019informed,nouwens2020dark,matte2020cookie,habib2020s,machuletz2020multiple}. This work investigates current implementations of CCPA requirements and their effects on user privacy, in particular their effect on users' awareness of and likelihood of invoking their right to opt-out of sale of their personal information.



As a first step, we performed a series of observational studies to understand how websites currently implement the opt-out of sale requirement imposed by CCPA. Our first observational study was a manual survey of the Alexa Top 500 U.S. websites conducted immediately after enforcement of CCPA began (between July 1-15, 2020); we reproduced the manual study in January 2021 on the same websites to understand how compliance has evolved over the first six months. We also performed an automated analysis of the top 5000 websites in January 2021.

Immediately after enforcement began in July 2020, we found that 207 (41.4\%) of the top 500 websites had some implementation of the user's right to opt-out of sale on the desktop version of their site and 116 (23.2\%) claimed that they did not sell personal information as defined by CCPA. 163 (33.5\%) pages both failed to provide an opt-out and did not specifically deny selling information as defined by CCPA, indicating potential non-compliance with the law. By January 2021, the number of if websites potentially not in compliance dropped to 55, with a slight increase in the number of sites that offered opt-out mechanisms (to 228) and a significant increase in sites that claimed to not sell personal information (to 207). However, our automated analysis (which did not categorize sites by whether they claimed to sell personal information) found that fewer websites in the Top 5000 provided CCPA-compliant opt-out links (21.7\%) than in the Top 500 (37.6\%).

Among the opt-out mechanisms observed in our manual observation studies, UI elements and design choices likely to decrease interaction were common.   Just 18 notified users of their right to opt-out of sale in banner, while most websites posted only the required homepage link, often visible only by scrolling down to the bottom of the site. Furthermore, even after clicking on a Do Not Sell link, many websites required significant additional work from users to opt-out, such as filling out forms or following directions for further steps, rather than presenting them with a single opt-out button. Finally, nudging (e.g., visually de-emphasizing the Do Not Sell button), unclear interfaces, and other \emph{dark patterns} were common.

To understand how the observed design choices affect users' behavior (e.g., whether they invoke their right to opt-out of sale) and users' awareness of their right to opt-out, we then conducted a pair of user studies. We implemented an aggregated news website that logged how users interacted with the website---particularly how they interacted with various different implementations of a Do Not Sell opt-out mechanism. We recruited 4357 unique California users through Google Ads and Amazon Mechanical Turk, and we observed their behavior. Users recruited through Amazon Mechanical Turk also completed a follow-up survey.

In our first experiment, we studied how the format of a Do Not Sell mechanism (link versus banner) affected user privacy. We found that users assigned to the link-only condition interacted with the Do Not Sell mechanism significantly less, and were significantly less likely to invoke their right to opt-out of sale compared to users who were shown a banner. Among Mechanical Turk users, those assigned to the link-only condition were significantly less likely to be aware that they had the right to opt-out of sale of their personal data than users assigned to conditions with banners. Among the conditions with banners, users were generally more likely to interact with banners located at the top of the page, particularly a full-width bar banner or a banner in the top right-hand corner.

In our second experiment, we studied the effect of nudging and inconvenience factors (e.g., having to fill out a form or having to select multiple options) on user interactions with the Do Not Sell mechanism. We found that highlighting did not significantly affect how users interact with the opt-out banner, but that other nudging mechanisms that appeared in the wild---including presenting the opt-out option as a link rather than a button or requiring users to click on a ``More Info'' link to access the opt-out mechanism---had a significant impact on how many users exercised their right to opt-out.  We also found that the inconvenience factors we considered all had significant negative effects on users' exercise of their right to opt-out. Conversely, we found that an anti-nudging design in which a banner contained only a single opt-out button (and no mechanism to explicitly accept the sale of personal information) significantly increased the user opt-out rate.

Overall, our findings suggest that compliance with the CCPA right to opt-out of sale is not yet universal, and that many companies who do comply with the law do so in ways that inhibit users from exercising their rights. We believe these results can serve as guidance for companies who want to enhance user privacy  by following the best-practices identified in this work, and we believe they should inform any future privacy regulations. 

\section{Background on CCPA}

The primary goal of the CCPA was to give users more control over their personal information. This resulted in the introduction of four key rights:
\begin{enumerate}
    \item \textbf{The right to know.} Users have a right to know what personal information a business collects and how that information is used and shared. This information should be communicated in a manner that provides the user with a ``meaningful understanding''. 
    \item \textbf{The right to delete.} Users have a right to delete personal information about them (with some exceptions).
    \item \textbf{The right to opt-out of sale.} Users have the right to opt-out of the sale of their personal information. Businesses must provide a ``a clear and conspicuous link'' on the homepage of their website entitled ``Do Not Sell My Personal Information'' that enables users to invoke their right to opt-out of sale. 
    \item \textbf{The right to non-discrimination.} Businesses cannot deny a service, degrade the quality of service, or change the price of a service just because a user exercises their rights under CCPA. 
\end{enumerate}
CCPA also broadened the definition of personal information to include any information ``that identifies, relates to, describes, is reasonably capable of being associated with, or could reasonably be linked, directly or indirectly, with a particular consumer or household''. This definition explicitly includes information about online activities (e.g., a user's interactions with a website) and any inferences drawn from personal information.

This work investigated how businesses implement the right to opt-out of sale, and evaluates how the design choices implemented by these businesses affect users' awareness of their right to opt-out of sale and their likelihood of invoking that right.

\section{CCPA Consent Notices in the Wild}

To investigate how websites implement the CCPA right to opt out of sale, we conducted a series of three observational studies. The first two studies were comprised of a manual analysis of the Alexa Top 500 U.S. websites (as listed July 1, 2020); the first manual study was conducted immediately after enforcement of CCPA began (July 1-15, 2020) and the second manual study was conducted six months later (January 1-February 26, 2021). In each manual study, we collected information about whether the website included an opt-out of sale link as well as information information about whether the website sold personal information (as defined under CCPA) and details about the design and implementation of the opt-out mechanism (if provided). We also implemented a crawler that visited each of the top 5000 U.S. websites (as listed on January 12, 2021) and automatically detected whether each website included an opt-out of sale link. The complete results of our observational studies are given in Table~\ref{table:ccpa_wild}.

\begin{table*}[t!]
\begin{center}
\begin{tabular}{ l rl rl rl rl}
\hline
\textbf{Sale of Data} & \multicolumn{2}{c}{\textbf{Top 500 (07/2020)}} &  \multicolumn{2}{c}{\textbf{Top 500 (01/2021)}} & \multicolumn{2}{c}{\textbf{Top 500  Mobile (07/2020)}}& \multicolumn{2}{c}{\textbf{Top 5000 (01/2021)}}\\
Sells Data & 207 & (41.4\%) & 228 &(45.6\%) & 205 & (41.0\%) & \multicolumn{2}{c}{-} \\  
Unspecified & 163 &(32.6\%) & 55 &(11\%) & 163&(32.5\%) & 4682& (93.6\%)\\
Does not Sell Data & 116 & (23.2\%) & 207 & (41.4\%) & 116& (23.2\%) & \multicolumn{2}{c}{-}\\
No info & 14 & (2.8\%) & 11 & (2.0\%) & 16 & (3.2\%) & 318& (6.4\%)\\

& & &&\\
\textbf{Opt-out of Sale} & \multicolumn{2}{c}{\textbf{Top 500 (07/2020)}} & \multicolumn{2}{c}{\textbf{Top 500  (01/2021)}} & \multicolumn{2}{c}{\textbf{Top 500 Mobile (07/2020)}} & \multicolumn{2}{c}{\textbf{Top 5000 (01/2021)}}\\
Banner & 18 & (3.7\%) & 18 & (3.7\%)  & 16 & (3.3\%) & \multicolumn{2}{c}{-} \\
\hspace{14pt} Of sells data & 18 & (8.7\%) & 18 & (7.9\%) & 16 & (7.8\%) & \multicolumn{2}{c}{-} \\
Valid Link on Homepage & \multicolumn{2}{c}{-} & 184 & (37.6\%) & \multicolumn{2}{c}{-}  & 1016 & (21.7\%) \\
\hspace{14pt} Of sells data & \multicolumn{2}{c}{-} & 182 & (79.8\%) & \multicolumn{2}{c}{-} & \multicolumn{2}{c}{-} \\
Any Link on Homepage & 174 & (35.8\%) & 196 & (40.1\%) & 169 & (34.9\%) & 1143&  (24.4\%) \\
\hspace{14pt} Of sells data & 173 & (83.6\%) & 195 & (85.6\%) & 166 & (81.0\%) & \multicolumn{2}{c}{-} \\
Any opt-out Mechanism & 207 & (42.6\%) & 228 &  (46.6\%) & 200 & (41.3\%) & \multicolumn{2}{c}{-} \\
\hspace{14pt} Of sells data & 201 & (97.1\%) & 226 & (99.1\%) & 192 & (93.6\%) & \multicolumn{2}{c}{-}  \\
No opt-out Mechanism & 279 & (57.4\%) & 261 & (53.4\%) & 284 &  (58.7\%) & 3539&  (75.6\%) \\
\hspace{14pt} Of sells data & 6 & (2.9\%) & 2 & (0.9\%) & 13 & (6.3\%) & \multicolumn{2}{c}{-} \\

\hline
\end{tabular}
\end{center}
\caption{CCPA Compliance in the Wild}\label{table:ccpa_wild}
\end{table*}

\subsection{Methodology}

Prior to beginning our manual observational studies, we developed a coding book based on a preliminary analysis of 50 websites (desktop and mobile); these codes included presence/absence of opt-out link (and visibility without scrolling), presence/absence of banner (and location), format of the opt-out mechanism (buttons, sliders, form, etc.), number of options in the opt-out mechanism, and presence and type of nudging. 

For each of the manual observational studies, one author visited each of the top 500 U.S. websites (as listed by Alexa on July 1, 2020) from a California IP address; for the first study, we also visited the site's mobile version from a smartphone. For each website, we recorded whether we were able to collected data about the website (i.e., whether the site returned a valid, English-language html page), whether the website claimed to sell user data as defined under CCPA, and whether the website contained an opt-out of sale link on their home page. If the website claimed to sell user data and did not have an opt-out of sale link on their home page, we checked whether there was an opt-out mechanism available in their privacy policy. If the website claimed to sell user data and had an opt-out mechanism, we qualitatively coded elements of the opt-out's design, such as how it was displayed and whether it employed nudging. Borderline cases were resolved through discussion with a second author.

To determine whether each website sold user information as defined under CCPA, we scanned through its privacy policy.  We looked for a specific statement regarding the sale of user data with regard to CCPA, as we noticed many websites made general statements denying data sale, only to later admit in a CCPA-specific section to engaging in sales as defined by the law. Websites that made no CCPA-specific statement about sale of data were coded as ``unspecified''.  

To determine whether websites had an opt-out of sale link on their home page, we performed a thorough manual check of each site's homepage. First, we searched each homepage for CCPA-related phrasing (``sell", ``info", ``CCPA", ``California", and ``CA"). We also manually checked any parts of the site where links were present, including any expandable page elements.  If the website claimed to sell personal data but did not have an opt-out link on the home page, we proceeded to search the site's privacy policy and, if applicable, CCPA umbrella page for instructions related to sale opt-out.

For websites where an opt-out mechanism was present, we additionally performed qualitative coding of UI design elements in its implementation.  We developed a coding book based on a preliminary analysis of 50 websites (desktop and mobile).  These codes included presence/absence of opt-out link (and whether immediately visible on the page), presence/absence of banner (along with location and in-banner options), format of the opt-out mechanism (buttons, sliders, form, etc.), and presence and type of nudging.  Each website was then visited from a California IP address by one author and manually coded. 

For the automated observational study, we implemented a web crawler using Node JS. The crawler visited each website on the list of Top 5000 U.S. websites, as listed by Alexa on January 12, 2021. For each website, the crawler first tried to visit the domain as listed; if that failed, it tried again with the \texttt{www} subdomain. If both requests resulted in errors, the entry was recorded as an error. For each site that returned a valid html page, the crawler searched the page first for links with the wording required by CCPA. If no such link was found, it then searched for links with a variety of alternate phrases drawn from our experience during the manual analysis (``california privacy'', ``consumer privacy'', ``do not sell'', and ``my info''); websites that contained such a link were recorded as containing invalid links.

\subsection{Data Sale and Compliance}

While we were unable to definitively judge any website's compliance with CCPA, our results outline the evolving landscape of company responses to CCPA in the six months after enforcement began. In our first manual study, we found that 41.4\% of the top 500 websites explicitly acknowledged that they sold personal information as defined under CCPA, 23.2\% of websites 23.9\% of those websites specifically stated that they did not sell user personal information, and 23.2\% of websites made no definitive statement either way. The remaining 14 websites returned errors when we attempted to access them.\footnote{Upon subsequent inspection, we found that the majority of these sites returned errors because the website actually used a non-standard subdomain and did not respond to requests made to the domain listed by Alexa or to the \texttt{www} subdomain. Two of these sites appear to have been temporary sites serving malware or adware that had been taken down prior to the date we attempted to visit them.} We also observed that there were significant differences in companies' interpretations of what constitutes a sale under CCPA. Notably, some companies assert that providing data to third-party tracking or targeted advertising tools does not constitute a sale despite the current consensus that this sort of transaction does constitute a sale of personal information under CCPA.\footnote{Note that during our first manual study, regulatory guidelines from the California Office of Attorney General were still under development and did not take effect until August 14th, 2020.} 

\begin{figure}[t!]
    \centering
    \includegraphics[width=\columnwidth]{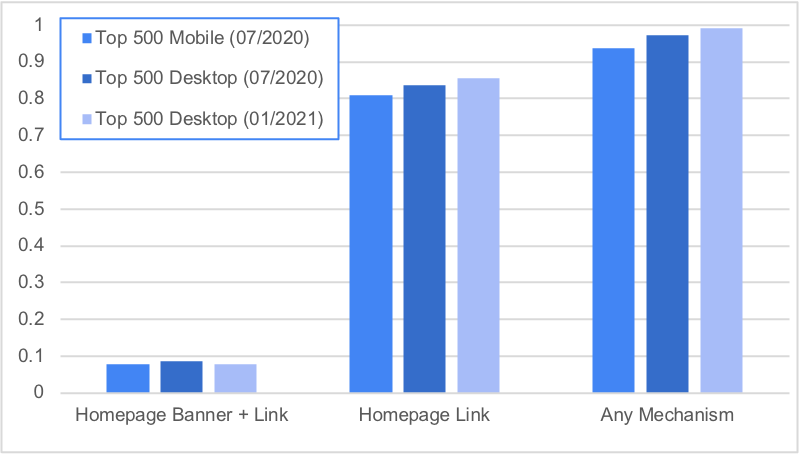}
    \caption{Prevalence of opt-out of sale mechanisms in Top 500 websites that sell personal information.}
    \label{fig:manual-links}
\end{figure}

In the second manual study---conducted six months after enforcement of CCPA commenced---we observed only 11\% of websites failed to specify whether or not they sold personal information, a significant decrease. Most of this change appeared to be due to websites that do not sell personal information now explicitly saying so (41.4\% of websites in the second study), although an additional 21 websites stated that they sold personal information in January 2021. These results suggest an increasing awareness of CCPA, and efforts to comply with its requirements, evolved over the six months after enforcement began.

To understand how websites implement the CCPA opt-out of sale requirement, we focused on the subset of websites that sell personal information, as defined under CCPA. In our first manual study, we found that 97.1\% of websites that explicitly acknowledge selling personal information provide some sort of opt-out of sale mechanism, however only 83.6\% of such websites provide on opt-out link on their home page as required by CCPA; compliance was slightly lower on the mobile versions of websites. We did observe improved compliance in our second manual study; by January 2021, 99.1\% of websites that were confirmed to sell personal information provided some sort of opt-out of sale mechanism, and 85.6\% provided an opt-out of sale link on their homepage (although only 79.8\% of such websites provided a link with the precise wording required by CCPA). These trends are summarized in Figure~\ref{fig:manual-links}. We observe, however, that actual compliance may be somewhat lower, as it is likely that some websites that do not specify their sale practices do sell personal information and therefore should be implementing the opt-out of sale requirement under CCPA.

\begin{figure}[t!]
    \centering
    \includegraphics[width=\columnwidth]{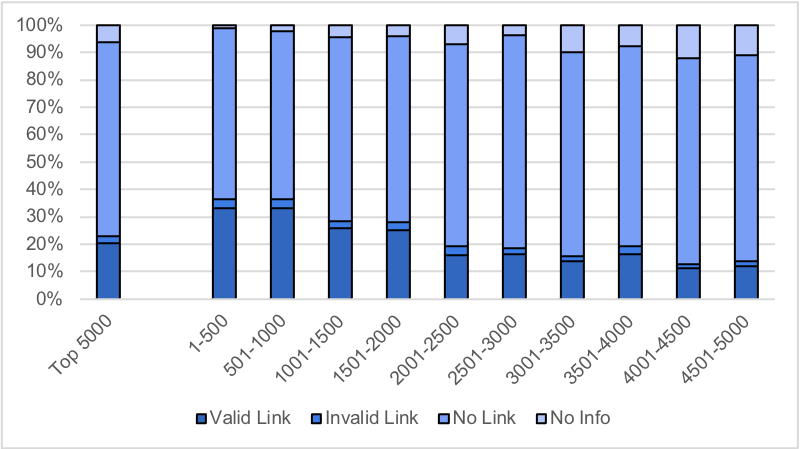}
    \caption{Prevalence of opt-out of sale links in Top 5000 websites.}
    \label{fig:automated-links}
\end{figure}

Our automated study extended our observations to the Top 5000 websites (as listed by Alexa in January 2021). However, our automated analysis was not able to classify sites according to whether or not that site sells user information, so we are not able to make definitive statements about rates of compliance with CCPA. Nonetheless, certain trends emerge. Our automated crawler found that 21.7\% of the Top 5000 websites visited had a valid opt-out link on their homepage compared to 34.2\% of the Top 500 websites visited.\footnote{The numbers for the automated analysis of the Top 500 websites differ slightly from the numbers for the January 2021 manual analysis because the two studies used top website lists from different dates.} An additional 3.5\% of websites in the Top 500 (and an additional 2.7\% of websites in the Top 5000) were found to contain a link that appeared related to CCPA but that did not use the legally-mandated language. As shown in Figure~\ref{fig:automated-links}, prevalence of opt-out links generally decreased among less frequently visited sites. We observe briefly that there are multiple possible explanations for this trend: it is possible that fewer lower ranked websites sell personal information or that more lower ranked websites are exempt from CCPA due to size. Nonetheless, our observations suggest that CCPA compliance is likely less pervasive among lower-ranked websites than among the Top 500 websites. 

\subsection{Form of Opt-out Mechanisms}

Among the Top 500 websites that implemented some form of opt-out of sale mechanism (207 desktop sites and 200 mobile sites in our July 2020 study and 228 desktop sites in our January 2021 study), we observed that websites commonly adopt design and implementation choices that appear to violate the spirit of the CCPA, which states that opt-out links should be ``clear and conspicuous'' on the homepage.

\subsubsection{CCPA Banners}

After the adoption of GDPR in the European Union, consent notices---banners that provide information about data collection practices and give the user an opportunity to consent---became pervasive~\cite{degeling2018we}. Substantial percentages of users are willing to engage with such notices, although factors such as position, options available, wording and nudging can mitigate the impact of such banners~\cite{utz2019informed}. We expected to see a similar rise in CCPA banners after enforcement began in July 2020, but our observational studies found that very few websites actually implemented opt-out mechanism in banner. In July 2020, just 18 of the Top 500 websites (16 on mobile) provided an opt-out mechanism in a banner; this number remained unchanged in January 2021. Moreover, many of the banners adopted design choices that have previously been found to decrease engagement~\cite{utz2019informed}: locating the banner as a bar along the bottom of the page or in the bottom right, allowing the user to interact with the page before interacting with the banner (no blocking), and linking to the opt-out mechanism rather than providing the mechanism directly in the banner. In one case, the banner was only visible after scrolling to the bottom of the page. Examples of each of the categories of banner ares shown in Figure~\ref{fig:banner-examples}, and precise statistics from each observation study are given in Table~\ref{table:banners-wild}.

\begin{table}[t!]
\begin{center}
\begin{tabular}{ l c c cc }
\hline
& \multicolumn{2}{c}{\textbf{07/2020}} &&\textbf{01/2021} \\
\cline{2-3}\cline{5-5}
 \textbf{Location} & \textbf{Desktop} & \textbf{Mobile} && \textbf{Desktop}\\
 Bottom bar & 77.8\% & 93.8\% && 72.2\% \\  
 Bottom left box & 0.0\% &- && 5.5\% \\
 Bottom center box & 5.5\% & - && 5.5\%\\
 Bottom right box & 11.1\% & - && 11.0\%\\
 Centered & 5.5\% & 6.3\% && 5.5\% \\
\\
& \multicolumn{2}{c}{\textbf{07/2020}} &&\textbf{01/2021} \\
\cline{2-3}\cline{5-5}
 \textbf{Options in Banner} & \textbf{Desktop} & \textbf{Mobile} && \textbf{Desktop}\\
Link to Mechanism & 83.3\% & 81.3\% && 77.7\% \\
 Single Do Not Sell button & 5.5\% & 6.3\% && 5.5\% \\
 Two or more buttons & 11.1\% & 12.5\% && 16.7\% \\  
  \\
  & \multicolumn{2}{c}{\textbf{07/2020}} &&\textbf{01/2021} \\
\cline{2-3}\cline{5-5}
 \textbf{Scrolling} & \textbf{Desktop} & \textbf{Mobile} && \textbf{Desktop}\\
 Visible without scrolling & 94.5\% & 100\%&& 94.5\% \\  
 No blocking & 5.5\% & 0\%  && 5.5\%\\
 \\
 & \multicolumn{2}{c}{\textbf{07/2020}} &&\textbf{01/2021} \\
\cline{2-3}\cline{5-5}
 \textbf{Blocking} & \textbf{Desktop} & \textbf{Mobile} && \textbf{Desktop}\\
 Blocking & 5.5\% & 6.3\%&& 5.5\% \\  
 No blocking & 94.4\% & 93.7 \%  && 94.4\%\\

 \hline
\end{tabular}
\end{center}
\caption{Properties of CCPA opt-out of sale banners}\label{table:banners-wild}
\end{table}

\subsubsection{Opt-out Links on Homepage}
CCPA mandates that opt-out of sale links must be ``clear and conspicous''. However, we found that many opt-out links were implemented in a manner likely to negatively impact usability.  Our findings are summarized in Table~\ref{table:links_wild}.

In July 2020, We found that 98.7\% of these links were located at the bottom of the page and that 97.4\% required scrolling down (often many screen-lengths) before they were visible, factors which contributed to users being unable to locate Do Not Sell links in a prior study~\cite{mahoney2020california}. The number of links that required scrolling was slightly lower in January 2021, but the difference was due to an increase in the number of hidden links (i.e., links that were visible only after opening a menu or otherwise interacting with the page); only 5 websites (2.7\%) had opt-out links that were visible without scrolling or clicking in January 2021. Moreover, links were often displayed among lists of other links and were typically in a smaller font size than the rest of the page. Some links were also displayed in low-contrast font colors (e.g., light gray). All of these factors have been found to impair link usability in other contexts~\cite{dev2020johnny}.

Mobile pages exacerbated many of the usability challenges we observed on desktop sites: mobile pages generally required significantly more scrolling to reach the opt-out link at the bottom of the page, and small font sizes appeared even smaller on a mobile phone screen. Furthermore, 19 (12.3\%) of opt-out links on mobile devices were hidden under one or more menu headings that would only expand to show the link after being clicked.  We recorded 17 different labels for these headings with no consistent pattern; many had names like ``Get Help'' or ``Explore'' that did not clearly denote that they contained the CCPA opt-out link.

\begin{table}[t!]
\begin{center}
\begin{tabular}{ l c c cc}
\hline
& \multicolumn{2}{c}{\textbf{07/2020}} &&\textbf{01/2021} \\
\cline{2-3}\cline{5-5}
 \textbf{Location} & \textbf{Desktop} & \textbf{Mobile} && \textbf{Desktop}\\
 Top & 1.28\% & 5.2\% && 0.6\% \\
 Left & 0.0\% & 0.6\% && 0.6\%\\
 Right & 0.0\% & 0.0\% && 0.6\%\\
 Bottom & 98.7\% & 94.2\% &&   98.3\% \\  

& & &&\\
& \multicolumn{2}{c}{\textbf{07/2020}} &&\textbf{01/2021} \\
\cline{2-3}\cline{5-5}
 \textbf{Scrolling} & \textbf{Desktop} & \textbf{Mobile} && \textbf{Desktop}\\
 Visible without scrolling & 2.6\% & 6.5\% && 7.2\%\\  
 Scrolling required to see & 97.4\% & 93.5\%  && 92.8\%\\
& & \\
& \multicolumn{2}{c}{\textbf{07/2020}} &&\textbf{01/2021} \\
\cline{2-3}\cline{5-5}
 \textbf{Visibility} & \textbf{Desktop} & \textbf{Mobile} && \textbf{Desktop}\\
 Hidden under clickable & 1.9\% & 12.3\% && 8.9\%\\  
 Visible without clicking & 98.1\% & 87.7\% && 91.1\% \\
\hline
\end{tabular}
\end{center}
\caption{Properties CCPA opt-out of sale links on homepages}\label{table:links_wild}
\end{table}

\begin{figure*}[t!]
    \centering
    \begin{subfigure}{.24\textwidth}
    \includegraphics[width=\textwidth]{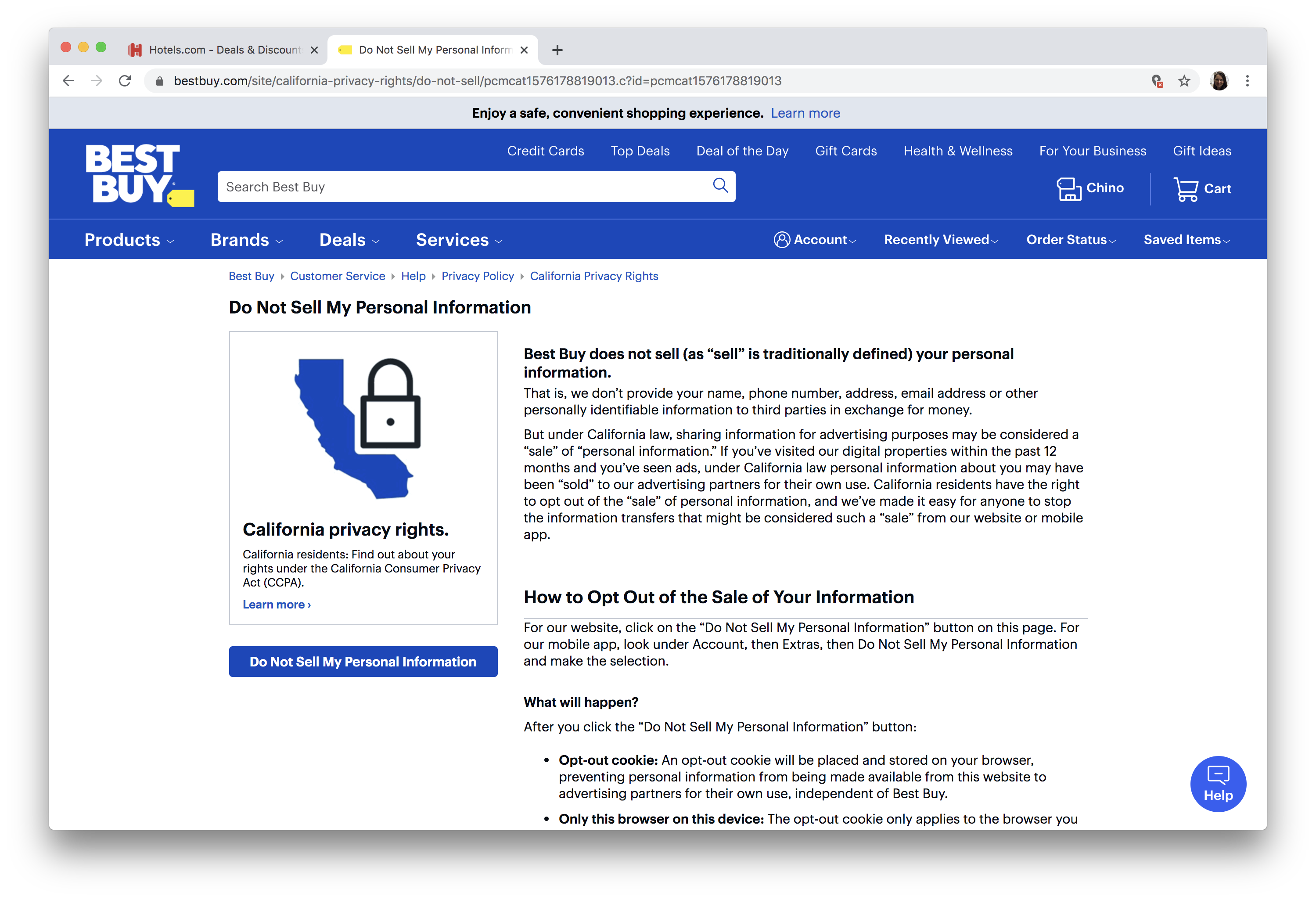}
    \caption{Example website with one option (Do Not Sell).}
    \end{subfigure}
    \begin{subfigure}{.24\textwidth}
    \includegraphics[width=\textwidth]{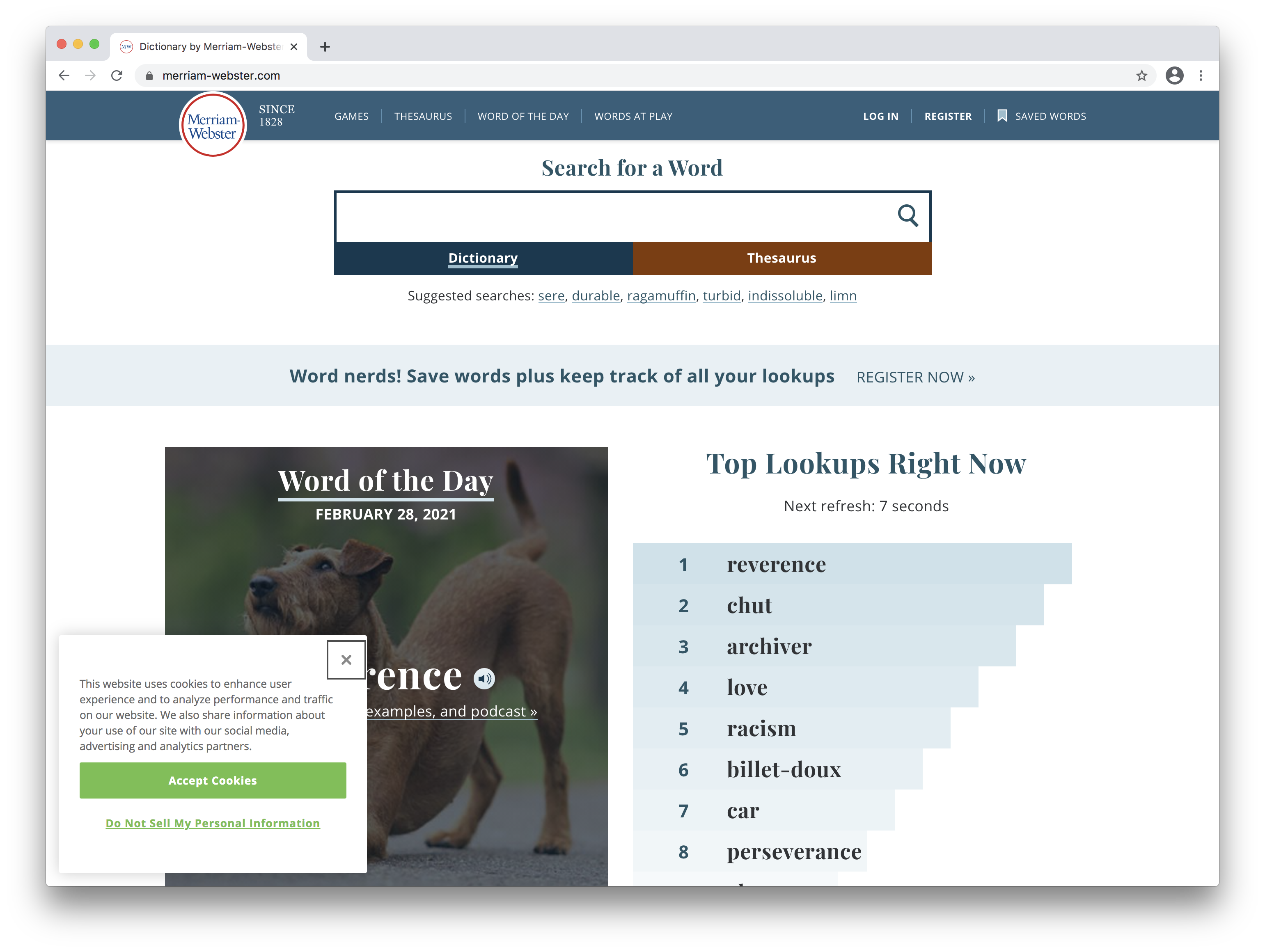}
    \caption{Example website with two options.}
    \end{subfigure}
    \begin{subfigure}{.24\textwidth}
    \includegraphics[width=\textwidth]{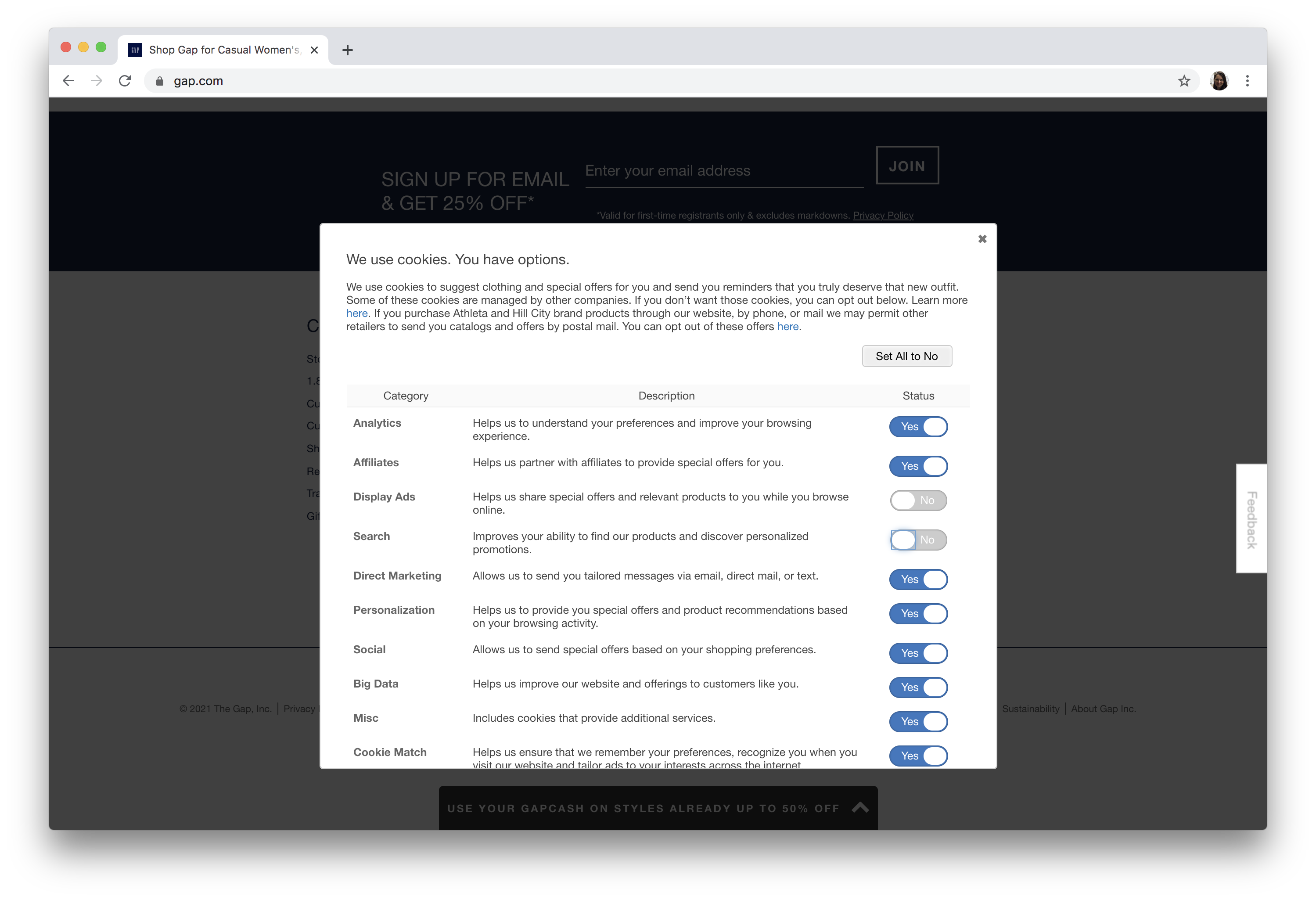}
    \caption{Example website with multi-option.}
    \end{subfigure}
    \begin{subfigure}{.24\textwidth}
    \includegraphics[width=\textwidth]{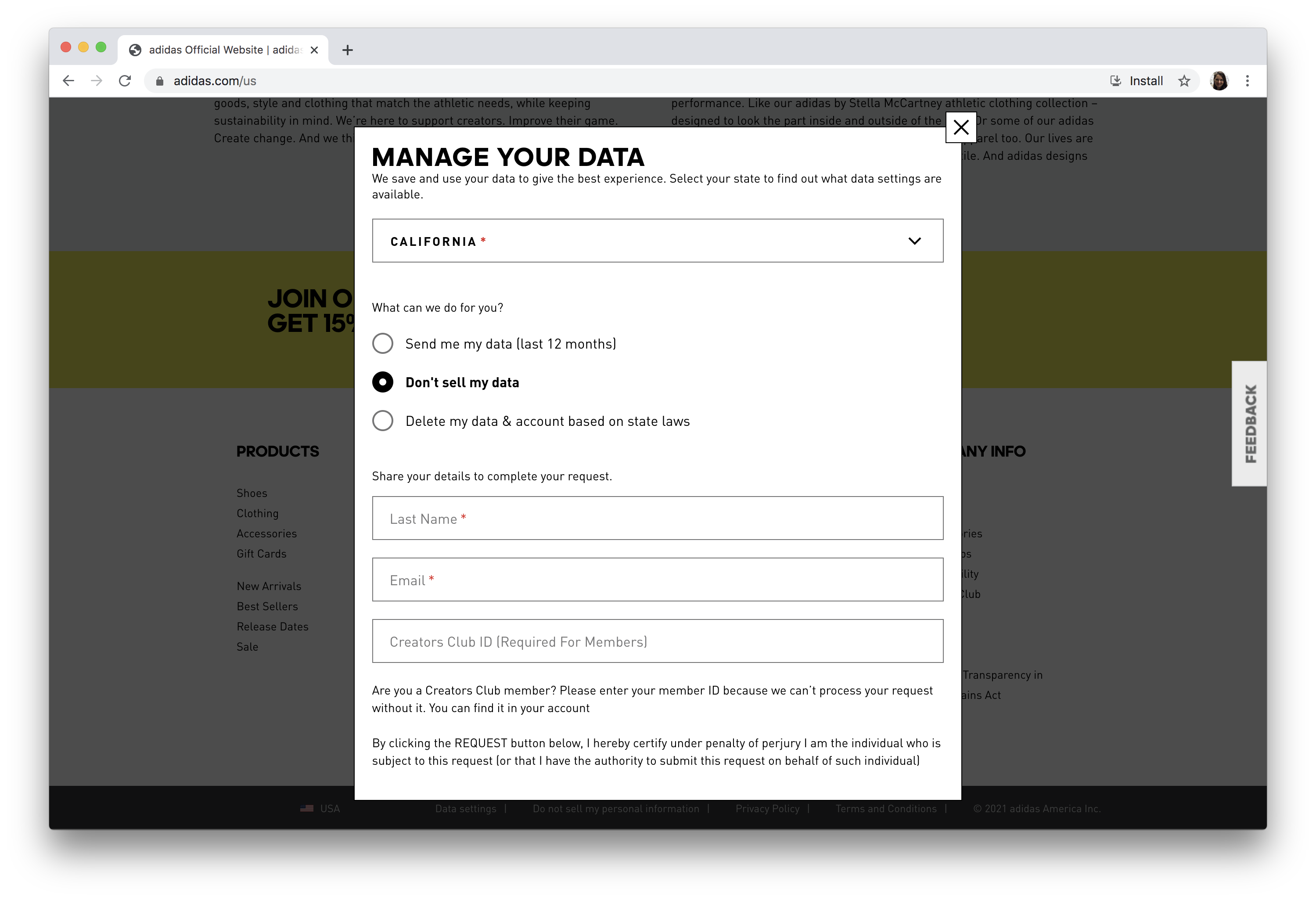}
    \caption{Example website fillable form.}
    \end{subfigure}
\caption{Example websites exhibiting different categories of opt-out choice implementations. The other four classes of implementation constitute written directions.}
    \label{fig:choices-examples}
\end{figure*} 

CCPA requires that opt-out links be titled ``Do Not Sell My Personal Information'' or ``Do Not Sell My Info''.  However, deviations from these mandated phrasings were common.  In July 2020, we documented 12 unique phrasings, most of which served to obfuscate the purpose of the link.  8 of these phrasings did so by omitting words (e.g., ``Do Not Sell''), which has been shown to be ineffective in communicating sale opt-out to users~\cite{cranor2020design}.  6 did so by adding technical or legal language to the link phrasing (e.g., ``Cal. Civ. Code §1798.135: Do Not Sell My Info'').  The importance of clear and consistent link names to enhance privacy choice usability has been demonstrated in other contexts~\cite{dev2020johnny, habib2020s, leon2012johnny, cranor2012can}, so these deviations seem likely to inhibit users' ability to utilize opt-outs. In January 2021, we categorized all websites according to whether their opt-out of sale link followed the legally-mandated wording; we found that 12 websites (6.7\%) used non-compliant language for their opt-out of sale link.

18 of the link-based pages (11.7\%) also displayed an unrelated privacy or cookie banner, which contained no information on CCPA.  These banners directed privacy-inclined users away from opt-out controls and instead towards general privacy policies or cookie settings.  Such presence of multiple privacy-related headings on a site has been shown to impair privacy choice usability as users struggle to select the correct page from a site's navigation menu~\cite{habib2020s}. These banners also frequently blocked the bottom of the screen in a way that made the opt-out link (usually located at the very bottom of the page) impossible to see without first dismissing the banner, in some cases even requiring users to make potentially unwanted privacy choices like accepting cookies in order to view the CCPA opt-out.  In a prior study, the presence of such a blocking cookie banner was noted to contribute to users being unable to find a site's DNS link when searching for it~\cite{mahoney2020california}.

\subsubsection{Opt-out Links in Privacy Policy}

In addition to the websites that provided an opt-out link on their homepage, a few websites (33 in the July 2020 study and 32 in the January 2021 study) provided an opt-out mechanism that was only accessible from the site's privacy policy. This implementation is not compliant with the requirements of CCPA---which states that an opt-out link must be provided on the homepage of the website. Moreover, prior work has consistently found that users do not read or look at privacy policies, so providing a mechanism that is only accessible from that page---and that is often embedded within long, legalese text---is likely to deter users from discovering or invoking their right to opt-out of the sale of their personal information.

\subsection{Opt-Out Controls}

Among both banners and links, the vast majority of homepage opt-out notices we observed (98.3\% in July 2020 and 98.0\% in January 2021) linked to a Do Not Sell page that required further action to opt-out.  We therefore also analyzed the design choices implemented by the opt-out controls that were reached after clicking on this link. We identified 8 classes of controls for opting out of sale of personal information:
\begin{enumerate}
\item \textbf{One option (Do Not Sell)}. A single clickable element that opts the user out of information sales.
\item \textbf{Two option}. A two-state control toggling opt-out entirely on or off.  This took the form of a toggle switch, a checkbox, or a pair of buttons (``Accept'' and ``Do Not Sell''). 
\item \textbf{Multi-option}.  Fine-grain options to control information sale beyond entirely enabled/disabled.  In most cases, these options allowed the author to authorize or disallow sale for different purposes of to different third-party companies. 
\item \textbf{Fillable form}. A form requiring the user to input personal information in order to opt out.  This information ranged from just an email address, to full name, address, and more.  These were frequently seen on shopping and subscription-based websites.
\item \textbf{Other directions for contacting company}. Instructions to contact the company (usually through email or a customer service portal) in order to make an opt-out request.
\item \textbf{Directions for dealing with other third party(s)}. Directions to opt out by using controls on third party websites, usually industry-provided targeted advertising opt-out sites.
\item \textbf{Directions for adjusting account settings}. Seen on websites where users have an account, these sites directed users to sign in and utilize privacy settings within their account to control sale of their information.
\item \textbf{Directions for adjusting browser settings}. Directions to opt-out of information sale by adjusting browser settings, usually either disabling cookies or enabling a “do not track” setting on mobile devices.
\end{enumerate}
The first three classes of mechanisms were classified as \emph{direct} mechanisms, since those mechanisms enable users to opt-out directly on that webpage; the other five classes of mechanisms, which gave instructions for adjusting settings, contacting the company, or contacting various third parties, were classified as indirect mechanisms.

\begin{figure}[t!]
    \centering
    \includegraphics[width=\columnwidth]{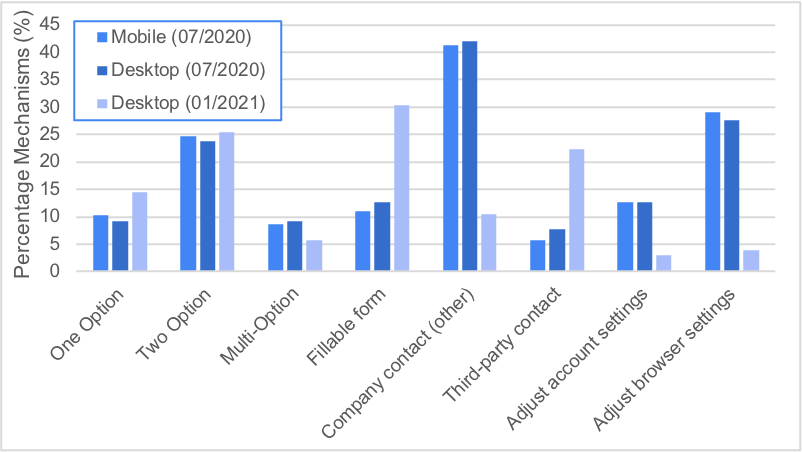}
    \caption{Prevalence of different CCPA opt-out controls in our observational studies. One Option, Two Option, and Multi-Option designs are considered to be direct mechanism. The other five classes of controls are indirect mechanisms. }
    \label{fig:optout-choices}
\end{figure}

Overall, 41.1\% of the Top 500 who provided an opt-out mechanisms in July 2020 provided some sort of direct mechanism; 9.2\% of sites provided a single Do Not Sell button, 23.7\% provided a two-option mechanism, and 9.2\% provided a multi-option mechanism. In January 2021, the fraction of opt-out mechanisms that provided some form of direct mechanism had risen slightly (to 46.9\%), and the fraction of mechanisms that provided a single Do Not Sell button had risen significantly (to 19.3\%). Although indirect mechanisms were the most common class of mechanism observed in both studies, we observed significant drops in the prevalence of fillable forms (from 42.0\% to 28.5\%) and in directions for adjusting browser settings (from 12.6\% to 3.5\%) between July 2020 and January 2021. The complete categorization of opt-out controls observed in our two manual observational studies is shown in Figure~\ref{fig:optout-choices}.  

\subsubsection{Nudging in Direct Mechanisms}

Most implementations of direct mechanisms offered two or more options; among these classes of direct mechanism, nudging---a class of design patterns intended to steer the user away from exercising their right to opt-out---was common.

When developing our nudging taxonomy, we looked for design Dark Patterns that had been previously identified as issues among GDPR consent banners.  Prior work has found default pre-selection, choice highlighting, and asymmetric UI all prevalent among cookie consent mechanisms; default pre-selection in particular is specifically banned by GDPR regulations~\cite{utz2019informed}.  Relatedly, we logged when mechanisms required unequal levels of difficulty to accept and opt-out from data sales, another subject of cookie banner research and regulation~\cite{nouwens2020dark}.  Finally, considered \emph{confirmshaming}, the practice of asking users to confirm decisions they have already made~\cite{brignull2019dark}.  Combining this prior work with the designs we found during our study, we documented the following categories of nudging:

\begin{enumerate}
\item \textbf{Default}. Switches, check boxes, or toggle controls used to opt-out were pre-selected to allow sale of personal information by default; the user needed to actively change the settings in order to opt-out. 
\item \textbf{Highlight}. Color was used to visually highlight the option for allowing information sale.
\item \textbf{Asymmetric UI}. The control to allow sale of information used a more visually prominent mechanism than the control to opt-out of sale (e.g., a clickable button vs. an inline link). 
\item \textbf{Confirmshaming}. A confirmation popped up after the user tried to opt-out,  and additional click(s) were required to finish the process.
\item \textbf{Asymmetric Difficulty}. Accepting sale of personal data was easier than opting out (e.g., immediately accessible with one click, while opt-out required toggling multiple options and/or visiting a separate opt-out page).
\end{enumerate}

\begin{figure}[t!]
    \centering
    \includegraphics[width=\columnwidth]{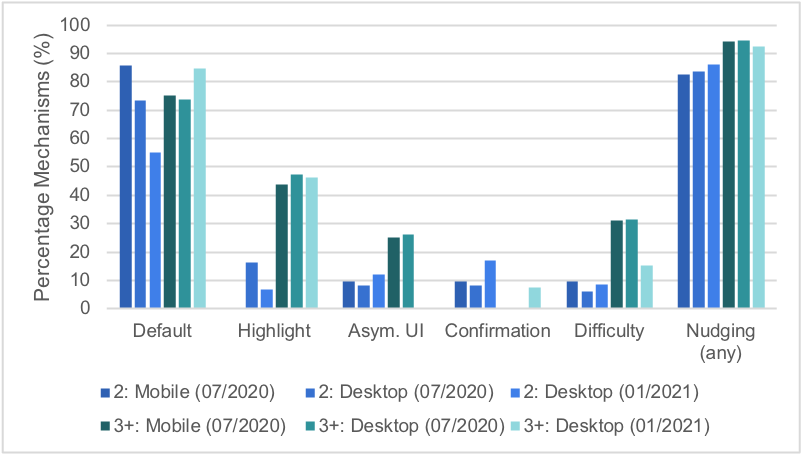}
    \caption{Prevalence of nudging in direct opt-out mechanisms with two or more options, by study. Data for mechanisms with 2 options are shown in blue, and data for mechanisms with 3 or more options are shown in gray. }
    \label{fig:optout-nudging}
\end{figure}

We found that nudging was prevalent among direct opt-out mechanisms: 73-85\% of two-option controls and 92-95\% of multi-option controls used some form of nudging. Defaulting was the most common form of nudging, but highlighting, difficulty, and asymmetric UI elements were also common in multi-option mechanisms.  These results are summarized in Figure~\ref{fig:optout-nudging}.

\subsubsection{Usability of Indirect Mechanisms}

The majority of opt-out mechanisms (57.9\% in July 2020 and 55.3\% in January 2021) were an instance of some class of indirect mechanism: directions for adjusting account settings, a fillable form or other instructions for contacting the company, or directions for dealing with browser settings or other third parties. Many of these implementations raise inherent usability concerns. 

Prior work has shown that users cannot or will not complete opt-outs on multiple third-party sites, suggesting directions for dealing with multiple third parties may be effectively unusable ~\cite{leon2012johnny}.  Even when users were required to utilize only one such site, the targeted ad opt-out sites to which users were directed have major documented usability issues~\cite{leon2012johnny, garlach2018m, habib2020s} and instructions were often lacking about how to opt-out once on a linked third-party site.  Online behavioral advertising opt-outs also do not fully opt the user out of tracking by third parties, only being shown targeted ads: half of third parties continue to track users after such opt-outs are set ~\cite{sakamoto2019after}. As such, relying on these pages as an opt out of for the sale of user information may not comply with CCPA, in that it does not actually allow users to stop the sale of their information.  Direction to one or more third parties was present in 27.6\% of opt-outs, making this a significant issue.

Directions to adjust account settings also may not comply with CCPA, which forbids companies from requiring users make an account to submit an opt-out.  Such directions furthermore often lacked specific information about which settings needed to be changed to opt out of sale, or how to proceed once signing in.  Directions for adjusting browser settings also generally lacked detailed instructions about how to perform specific changes required to opt out, instead directing users to general browser information pages about adjusting cookie settings.

Directions for contacting the company raised concerns due to past work showing the difficulties users have with such directions.  A prior study of Do Not Sell found that users contacting companies for an opt-out request are frequently met with slow responses, requests for invasive or difficult-to-provide information, and unclear instructions in company responses~\cite{mahoney2020california}.  Another study of other privacy opt-outs found that, in addition to adding to the number of required actions, directions for contacting the company left users struggling to know what information to include in such requests~\cite{habib2020s}.

Multi-step instructions generally were of such length and complexity to raise usability concerns.  These instructions typically contained ambiguous and poor directions that were missing key information, such as which steps were required to completely opt out.  Some contained contradictory instructions: for instance, one site told the user to both keep cookies enabled in order to store an opt-out cookie, and to disable cookies entirely.  Some also entailed individually visiting an extensive number of third party sites and independently navigating their opt-out processes (as many as 16 for one website).

Finally, fillable forms, the most common type of opt-out, at times requested excessive information from users.  While some user information may need to be obtained to correctly stop the sale of their information, we documented examples of information being requested that was obviously unnecessary to the opt-out process.  For instance, several pages required even non-account holding web users to input an email address by claiming it was necessary to create a ``verifiable" request, a requirement which other websites did not find necessary.  One page also required the the user's full physical address to confirm California residency.  Such practices both force additional work on users to complete the request and may raise concerns as to how the provided information will be secured or utilized, turning the opt-out process itself into a potential privacy threat.  For instance, at least one previously documented case exists of a company using an email address provided during CCPA sale opt-out to sign a user up for a marketing list, despite the law explicitly prohibiting such practice~\cite{mahoney2020california}.

\section{The Impact of CCPA Design Choices}

To evaluate the effect of the observed opt-out mechanisms on user privacy, we conducted a user study in which we observed how users interacted with various different opt-out of sale mechanisms. This study evaluated two primary research questions: (1) What is the impact of link-only mechanisms compared to banner mechanisms? and (2) What is the impact of nudging? We evaluated the impact of these design factors on how likely users were to opt-out of the sale of their personal information and the impact on how aware users were that they had a right to opt-out of sale. 

\subsection{Methodology}

\begin{figure}[t!]
    \centering
    \includegraphics[height=1.6in]{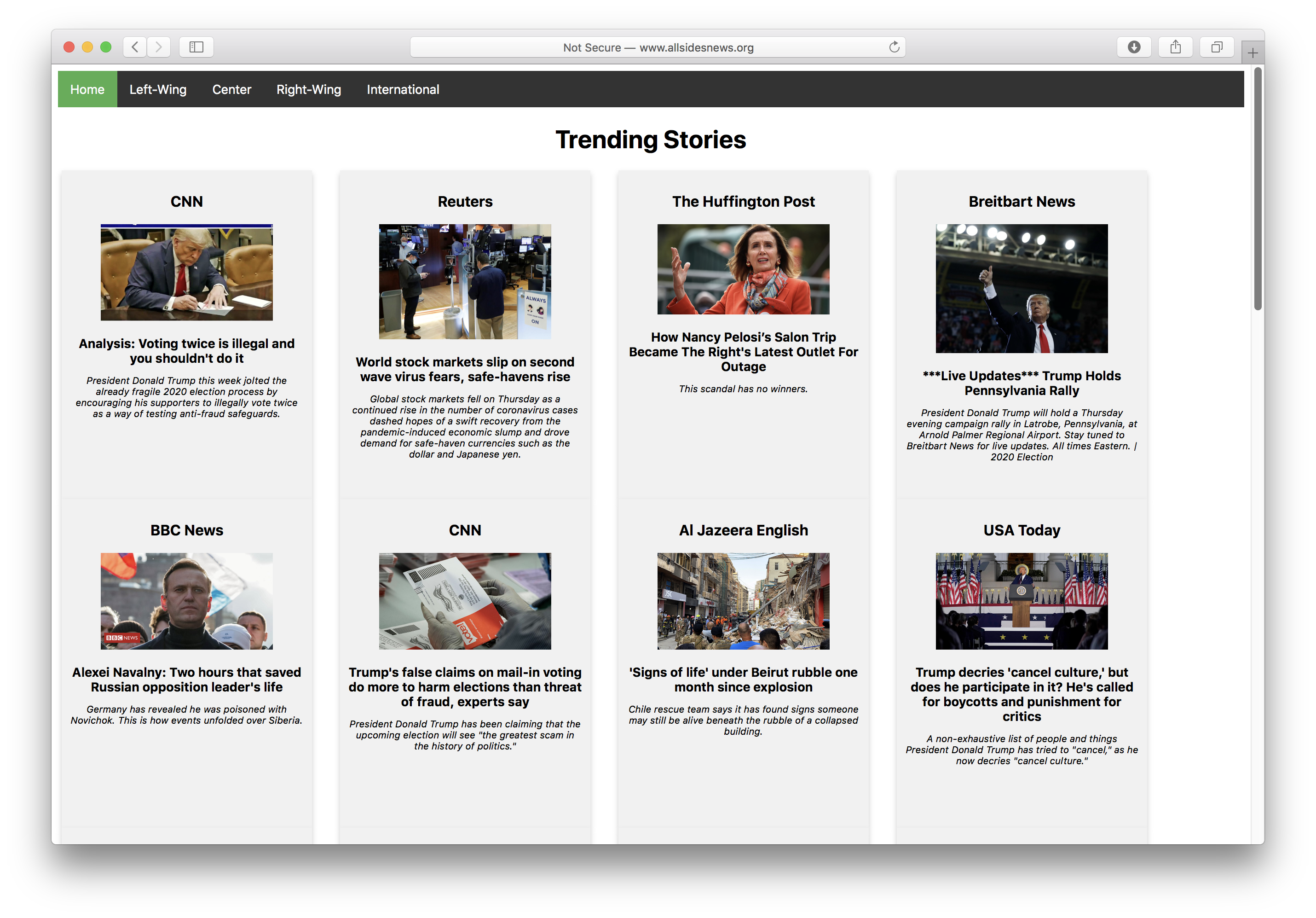}
    \caption{News aggregation site used in user studies.}
    \label{fig:website-screenshot}
\end{figure}

To conduct this study, we implemented a news aggregation site to serve as our example website.  We chose this context because it provides a credible privacy threat for even a brief, one-time visitor: a user's browsing pattern alone on such a site can expose sensitive information such as political beliefs and interests that are commonly sold to third party aggregators. A screenshot of our website in shown in Figure~\ref{fig:website-screenshot}. 
For each experiment, we implemented several different versions with different opt-out mechanisms (i.e., conditions); each study participant was pseudo-randomly assigned to a condition based on the current IP address. The details of the individual conditions are described in Sections~\ref{section:design-location} and \ref{section:design-nudging}.

For each user who visited the site, we logged how that user interacted with the site. Logged actions were associated with a unique identifier constructed by hashing the IP address of the user; no personally-identifiable information was stored. Logged actions included various ways of interacting with the opt-out mechanisms, including clicking on the opt-out link, clicking on individual buttons in a banner, or closing a banner. We also logged general interactions with the site: which pages the user visited and which links they clicked on. Finally, we logged a heartbeat message whenever the webpage was in focus on the user's device to enable us to calculate how long each user spent on the site. 

\begin{figure}[t!]
    \centering
    \includegraphics[width=\columnwidth]{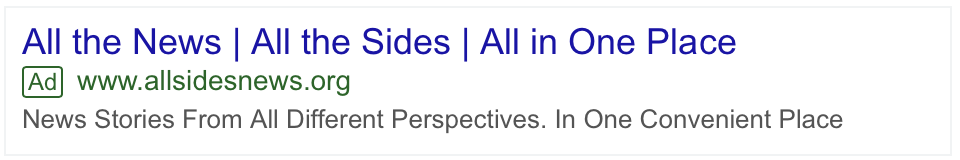}
    \caption{The Google Advertisement displayed to users.}
    \label{fig:google-ad}
\end{figure}  

At the bottom of each page (adjacent to the opt-out link) we provided a link to our privacy policy, which stated that this was an academic study exploring how users interact with Do Not Sell mechanisms, and that no personal information was collected; the privacy policy also included a button that users could click to opt-out of the study and have their log entries deleted.

Users were recruited by two methods: through Amazon Mechanical Turk and through Google Ads. To avoid biasing the results, none of the participants were informed upfront that they were participating in a study about CCPA or privacy. 

To recruit participants through Amazon Mechanical Turk, we advertised the task as, ``5 min - Beta Test Aggregated News Website”. Users were asked to visit our site and interact however they normally would with a webpage; they then completed a follow-up survey containing questions about the sale of data on the page, as well as demographic information (detailed in Appendix A).  Recruitment was limited to workers with at least a 95\% approval rate and at least 50 accepted HITs who were located in California.  Each worker was compensated \$1.00 USD for their participation, and the study ran from July 16-27, 2020.

One potential concern among MTurk users, however, was the possibility that their interactions may not accurately reflect the way users interact with CCPA mechanisms on real sites, due to their knowledge the site was part of a study and the paid motivation behind their visit.  To mitigate this concern, we recruited a second group of users through a Google Ads campaign run between August 15-18, 2020; the ad was placed for search terms relating to news, and was targeted at California users through the Google Ads network (see Figure) with an average cost per click of 33 cents. A copy of the recruiting ad is shown in Figure~\ref{fig:google-ad}. 

Our cleaned dataset included log records from 4357 unique users: 1726 users participated in Experiment 1 (1295 recruited through Google ads and 431 recruited through Amazon Mechanical Turk) and 2531 users participated in Experiment 2 (2233 recruited through Google ads and 398 recruited through Amazon Mechanical Turk). 53.3\% of our users visited the desktop version of the site and 47.7\% visited from a mobile browser.  

\begin{figure*}[t!]
    \begin{center}
    \begin{subfigure}[b]{0.3\textwidth}
        \centering
        \includegraphics[height=1.2in]{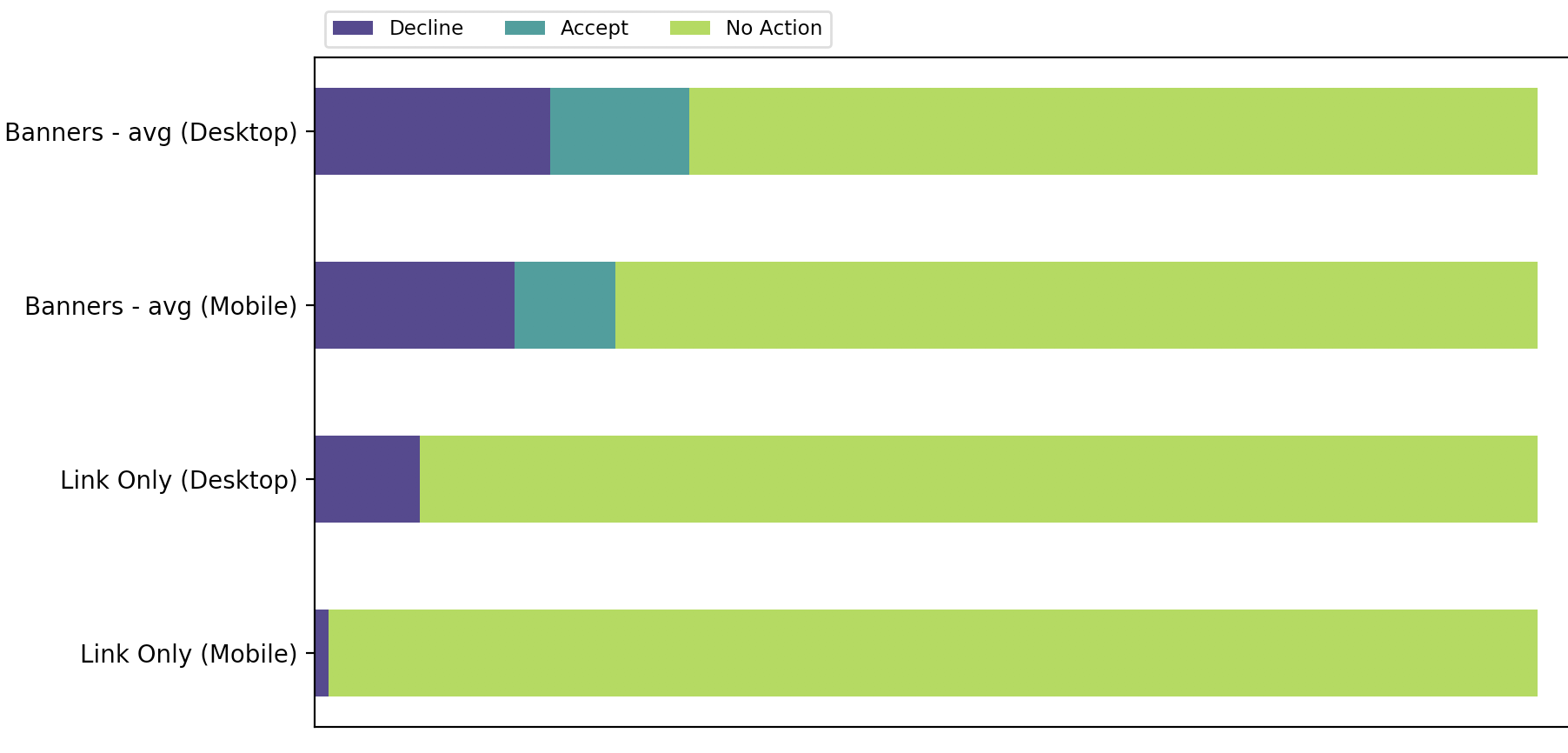}
        \caption{Format (banners vs. links)}
        \label{fig:interaction-location}
    \end{subfigure}
     \begin{subfigure}[b]{0.3\textwidth}
        \centering
        \includegraphics[height=1.2in]{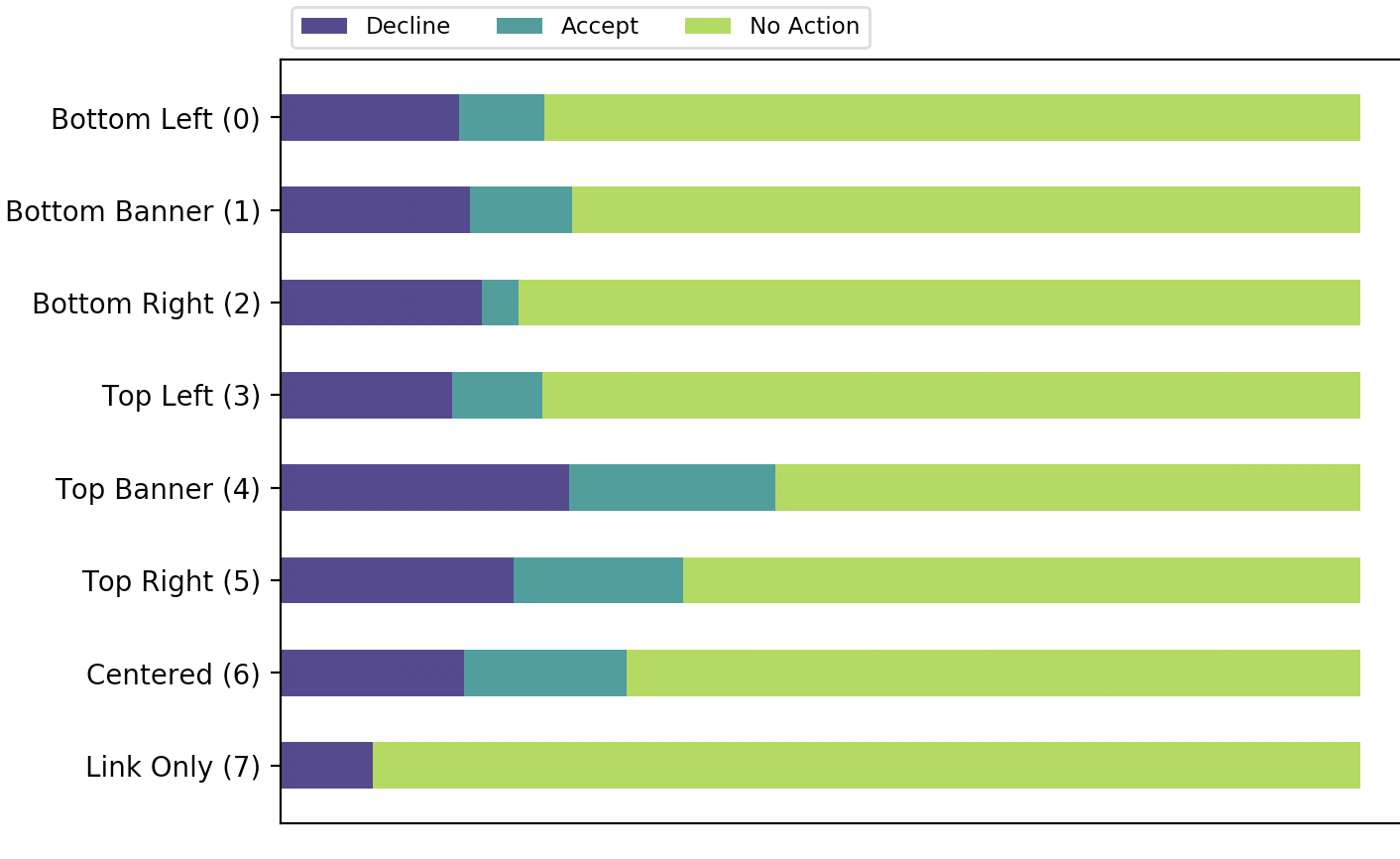}
        \caption{Desktop location}
        \label{fig:demographics-age}
    \end{subfigure}
    \begin{subfigure}[b]{0.3\textwidth}
        \centering
        \includegraphics[height=1.2in]{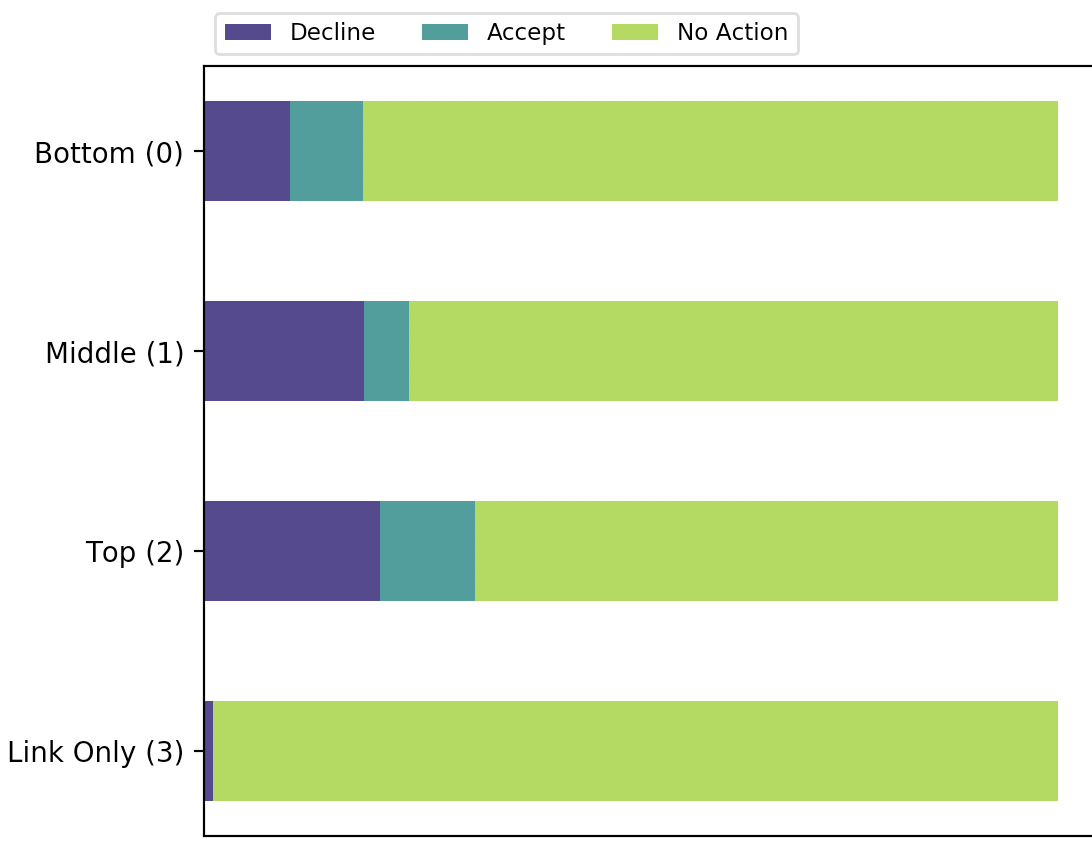}
        \caption{Mobile location}
        \label{fig:demographics-race}
    \end{subfigure}
    \end{center}

\caption{The effect of format (link vs. banner) and location on opt-out interaction rates. }\label{fig:interaction-location-banners}
\end{figure*}

\subsection{Analysis Plan}

The primary goal of our user study was to determine whether there were differences in how users interacted with the website differed based on the condition to which they were assigned. We therefore evaluated all hypotheses using Chi-squared contingency tests to test for differences in behavior and responses between study participants assigned to the different conditions. Due to the number of statistical tests, we applied a Holm-Bonferroni multiple comparison correction; all results that are reported as statistically significant are significant after the correction was applied.

\subsection{Ethical Considerations}

The majority of our study participants (3528/4357) were users recruited through Google Ads. To assure data validity, these users were not aware they were participating in a research study; this omission of prior informed consent inherently raises ethical issues. 

To minimize the risk to our users, no personally-identifiable information was collected. Log entries were associated with a unique identifier defined by a hash of the user's IP address; no IP addresses or other identifiers were stored. Information collected was used only for research purposes; no information was actually sold. 

We sought consent (and provided an option to opt-out) to the maximum extent possible. The privacy policy for our website clearly stated that this was an academic study exploring how users interact with Do Not Sell mechanisms, and that no personal information was collected; the privacy policy also included a button that users could click to opt-out of the study and have their log entries deleted. Users recruited from Amazon Mechanical Turk were informed of our data collection practices in advance and were given the opportunity to consent or to opt-out prior to beginning the user study.

This research received a waiver from the institutional ethics review board (IRB) at our institution.

\begin{figure}[t!]
    \centering
    \begin{subfigure}{2in}
    \includegraphics[height=1.6in]{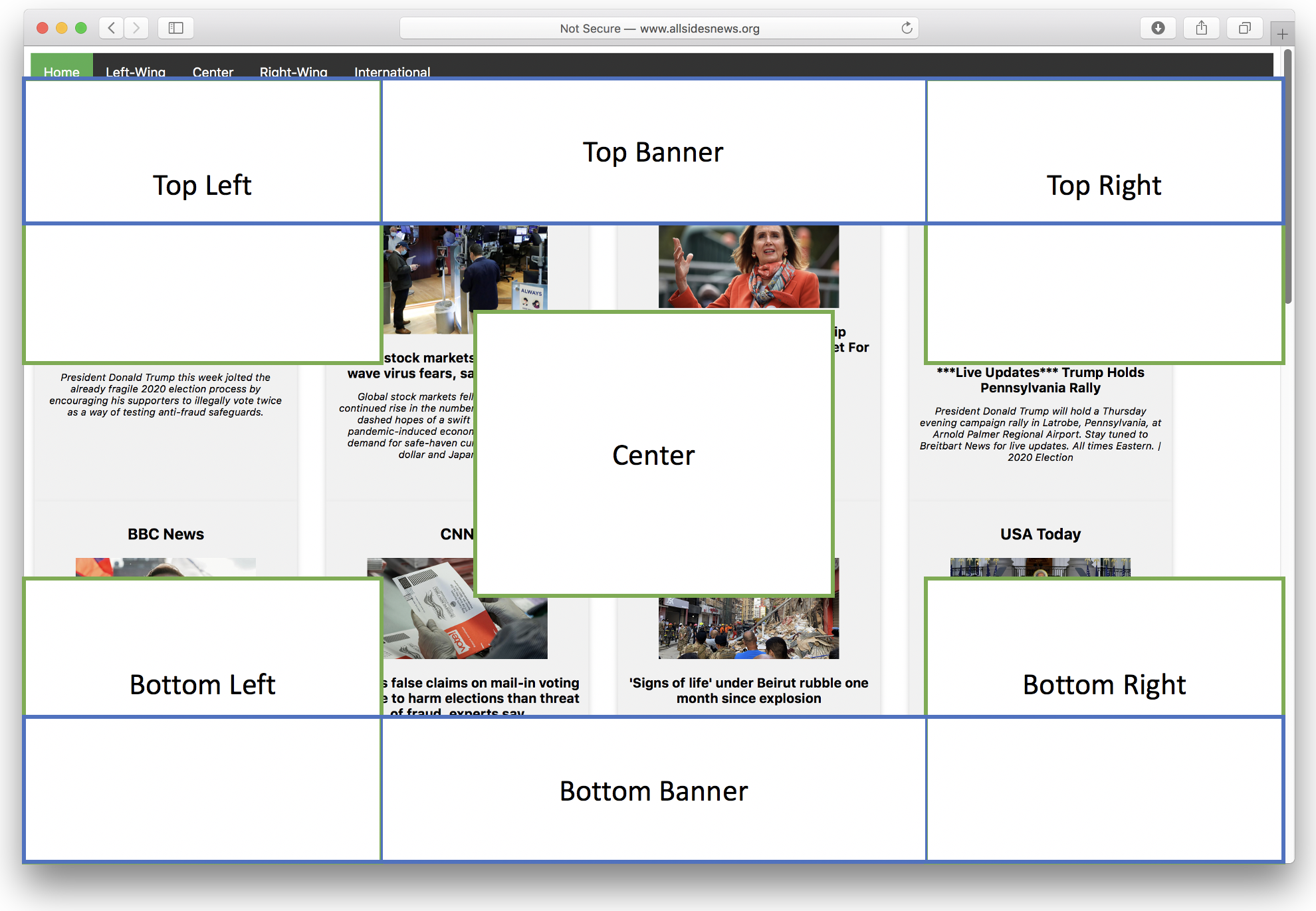}
    \caption{Desktop}
    \end{subfigure}
    \hspace{20pt}
    \begin{subfigure}{1in}
    \centering
    \includegraphics[height=1.6in]{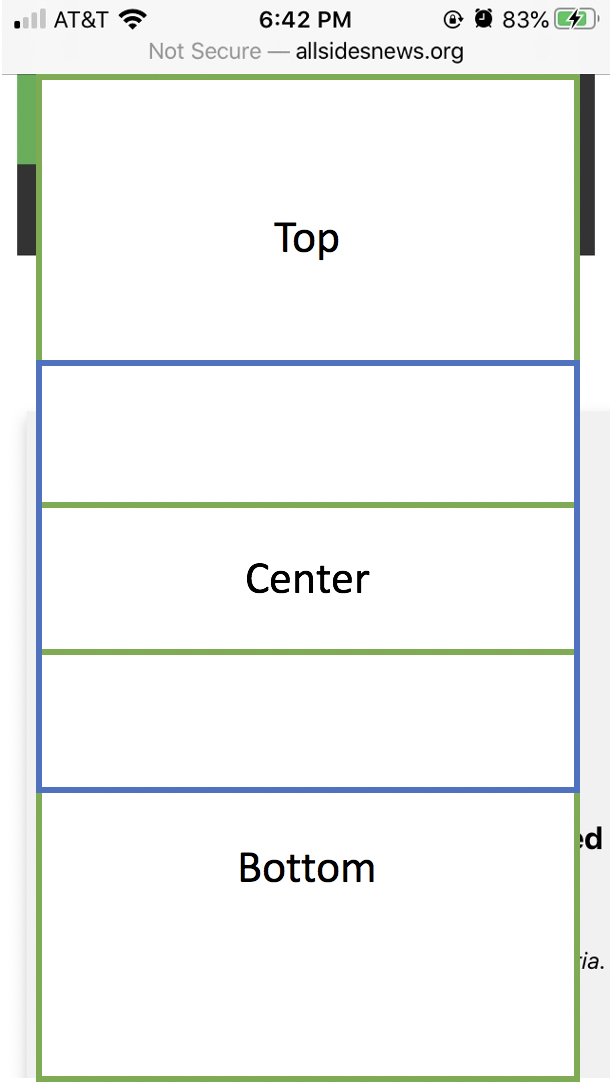}
    \caption{Mobile}
    \end{subfigure}
        \caption{A schematic diagram of the banner locations used in Experiment 1}\label{fig:location-diagram}
\end{figure}

\subsection{Experiment 1: Links vs. Banners}\label{section:design-location}

\begin{figure}[t!]
     \begin{subfigure}[b]{.5\textwidth}
     \begin{center}
        \includegraphics[height=1.0in]{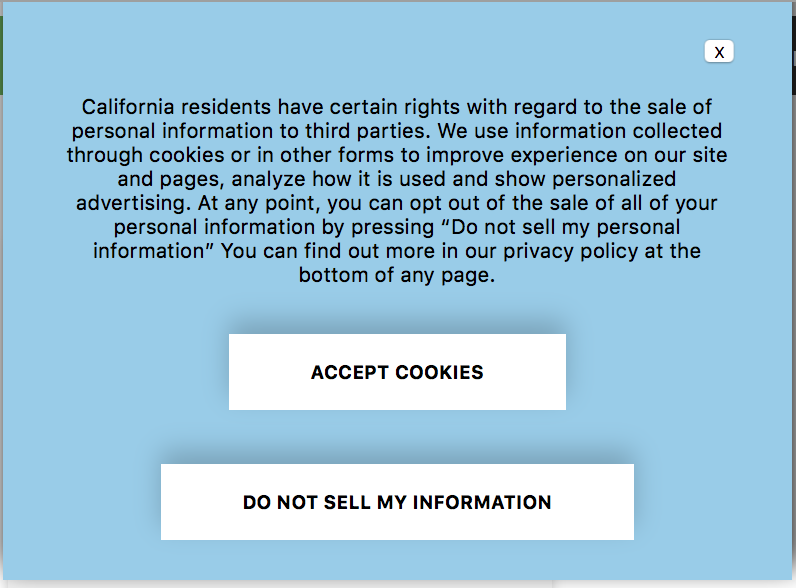}

        \caption{Pop-up banner used in the four corner locations and the center location (on desktop) and in all banner locations on mobile}
    \end{center}
    \end{subfigure}\\
    \begin{subfigure}[b]{.5\textwidth}
        \includegraphics[width=\textwidth]{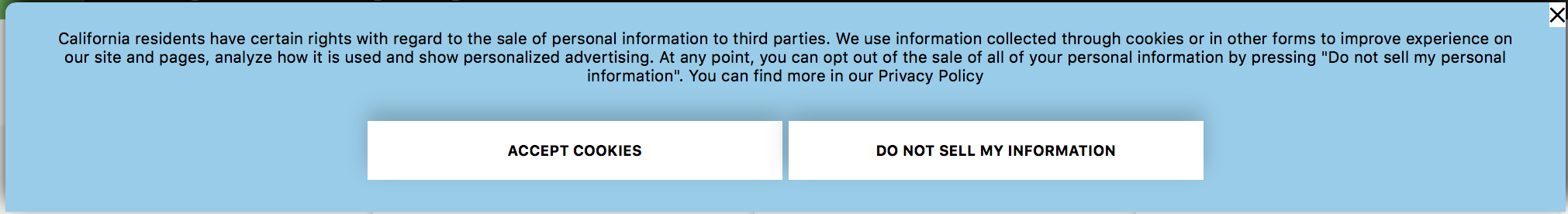}
        \caption{Full-width banner used in the Top Banner and Bottom Banner conditions on desktop}
    \end{subfigure}
\caption{Banner designs used in Experiment 1}\label{fig:location-banners}
\end{figure}

Our first experiment investigated how users interact with opt-out links compared to how users interact with opt-out mechanisms contained in pop-up banners in a between-subjects study. Prior work has found that user actions with banners vary significantly depending on the location of the banner~\cite{utz2019informed,nouwens2020dark,cantoni2013banner,doyle1997banner}, so we considered a variety of different banner locations. On the desktop version of the site, we used seven different banner locations: a pop-up banner in each of the four corners, a pop-up banner in the center of the page, and full-width banners at the top and bottom of the page. On the mobile version, we used three different banner locations: top, center, and bottom.  Each subject was assigned one condition at random when they first visited the page, with 106-188 participants assigned to each condition. A diagram of the different banner locations is provided in Figure~\ref{fig:location-diagram}; screenshots of the banners used in this experiment are provided in Figure~\ref{fig:location-banners}.

Unsurprisingly, users interacted with the opt-out link significantly less frequently than with banners ($p<.001$). On desktop, an average of 19.3\% of users who were shown a banner opted out of sale; for users who were shown a version of the website that only included the opt-out link at the bottom of the page, only 8.6\% of users opted out of the sale of their data. Overall interactions on mobile were lower, but we still saw significantly less interactions on the link-only design than on the designs with banners. On mobile, an average of 16.3\% of users opted out of the sale of their data compared to just 1.1\% of users who were shown the link-only version of the page ($p < .001$). These results are summarized in Figure~\ref{fig:interaction-location}. Moreover, interaction rates with the link-only condition were even lower among participants who clicked on a Google ad (who were unaware that they were participating in a study) than among users who were recruited through Mechanical Turk (who believed they were beta testing the website), so we suspect that the engagement gap between links and banners would be even larger for a real website.

We also evaluated the effect of banner placement on interaction rates; we found that users interacted most frequently with banners located at the top of the page or in the center of the page compared to banners located at the bottom of the page. The banner that ran across the full width of the top of the page exhibited the highest interaction rates: 26.7\% of users exercised their right to opt-out and 19.1\% explicitly accepted the sale of their personal information. This rate of interaction was significantly higher than each of the bottom locations ($p < .001$, $p = .005$, $p < .001$) and than the top-left location ($p = .002$), but it was not significantly higher than the other top locations. All banner locations individually resulted in significantly higher interaction rates that the link-only design ($p=.006$ for the bottom right location, $p < .001$ for all other locations). These results are summarized in Figure~\ref{fig:interaction-location-banners}.

To further understand the effect of opt-out links on privacy, we compared follow-up survey responses from MTurk users to the condition they were assigned. 
We found that only 54.5\% of users assigned to the link-only case were aware that they had the ability to opt-out of the sale of their personal data on our website compared to 71.1\% of users who were assigned to one of the conditions with banners ($p = .019$). This suggests that the predominant link-only implementation of CCPA's right to opt out of sale is significantly less effective at informing users of their rights than banner designs. 


\subsection{Experiment 2: Nudging}\label{section:design-nudging}

\begin{figure}[t!]
     \begin{subfigure}[b]{.5\textwidth}
        \includegraphics[width=\textwidth]{figures/nudging-baseline.png}
        \caption{Baseline: Neutral}
    \end{subfigure}\\
    \begin{subfigure}[b]{.5\textwidth}
        \includegraphics[width=\textwidth]{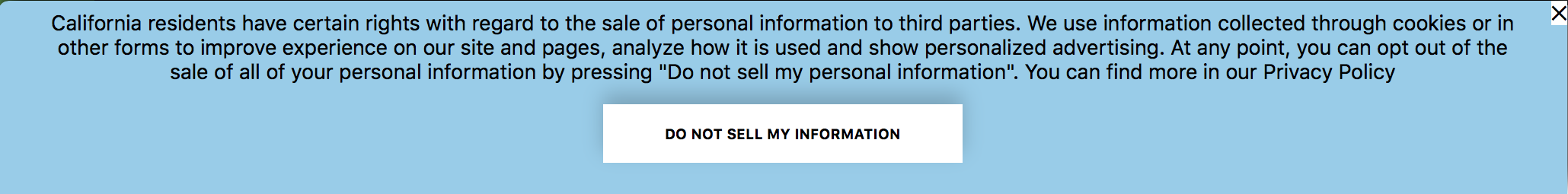}
        \caption{Anti-Nudging: Opt-out only}
    \end{subfigure}\\
    \begin{subfigure}[b]{.5\textwidth}
        \includegraphics[width=\textwidth]{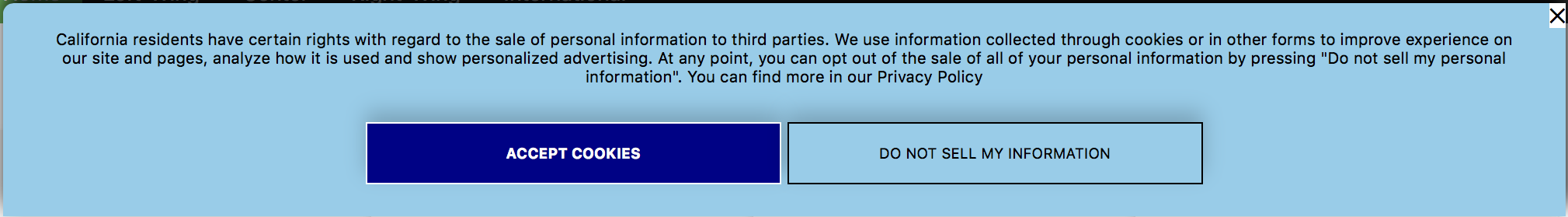}
        \caption{Nudging: Highlighting}
    \end{subfigure}\\
    \begin{subfigure}[b]{.5\textwidth}
        \includegraphics[width=\textwidth]{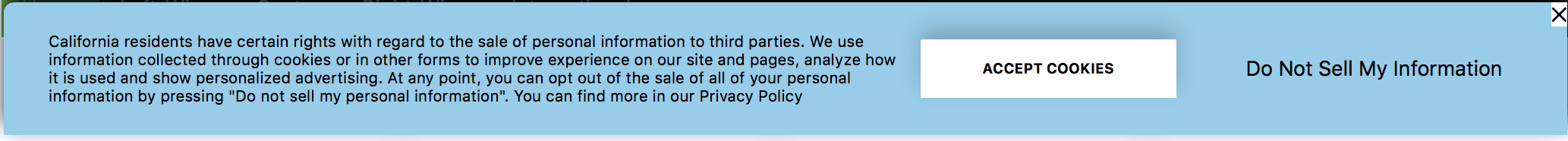}
        \caption{Nudging: Opt-out Link}
    \end{subfigure}\\
    \begin{subfigure}[b]{.5\textwidth}
        \includegraphics[width=\textwidth]{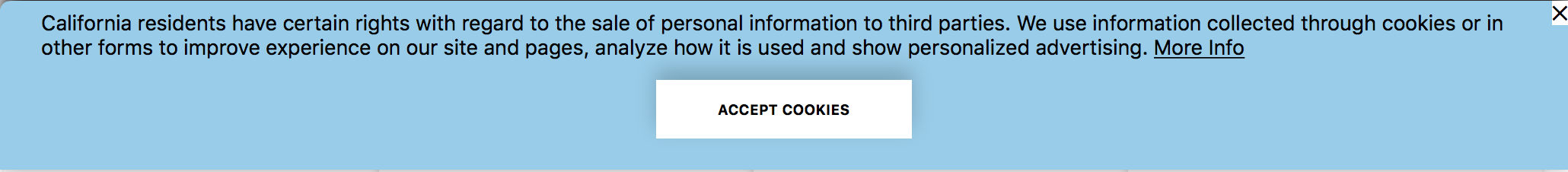}
        \caption{Nudging: Opt-out Inlined Link}
    \end{subfigure}
\caption{Banner designs used in Experiment 2}\label{fig:nudging-conditions}
\end{figure}

In our second experiment, we evaluated the effect of nudging and inconvenience factors on user interactions with the Do Not Sell mechanism. Drawing on inspiration from example implementations we saw in the wild, we generated three different banner designs with nudging: one in which the ``Accept'' button was highlighted (and the opt-out button was the same color as the background), one in which the opt-out button was replaced with a link, and one in which the opt-out button was replaced with a ``More info'' link inlined with the text. We compared these designs to a neutral design with no nudging and to an anti-nudging design in which the banner contained only a single opt-out button. To maximize data collection, all banners were implemented as full-width banners located at the top of the page; these banner designs are shown in Figure~\ref{fig:nudging-conditions}.

\begin{figure}[t!]
     \centering
        \includegraphics[width=\columnwidth]{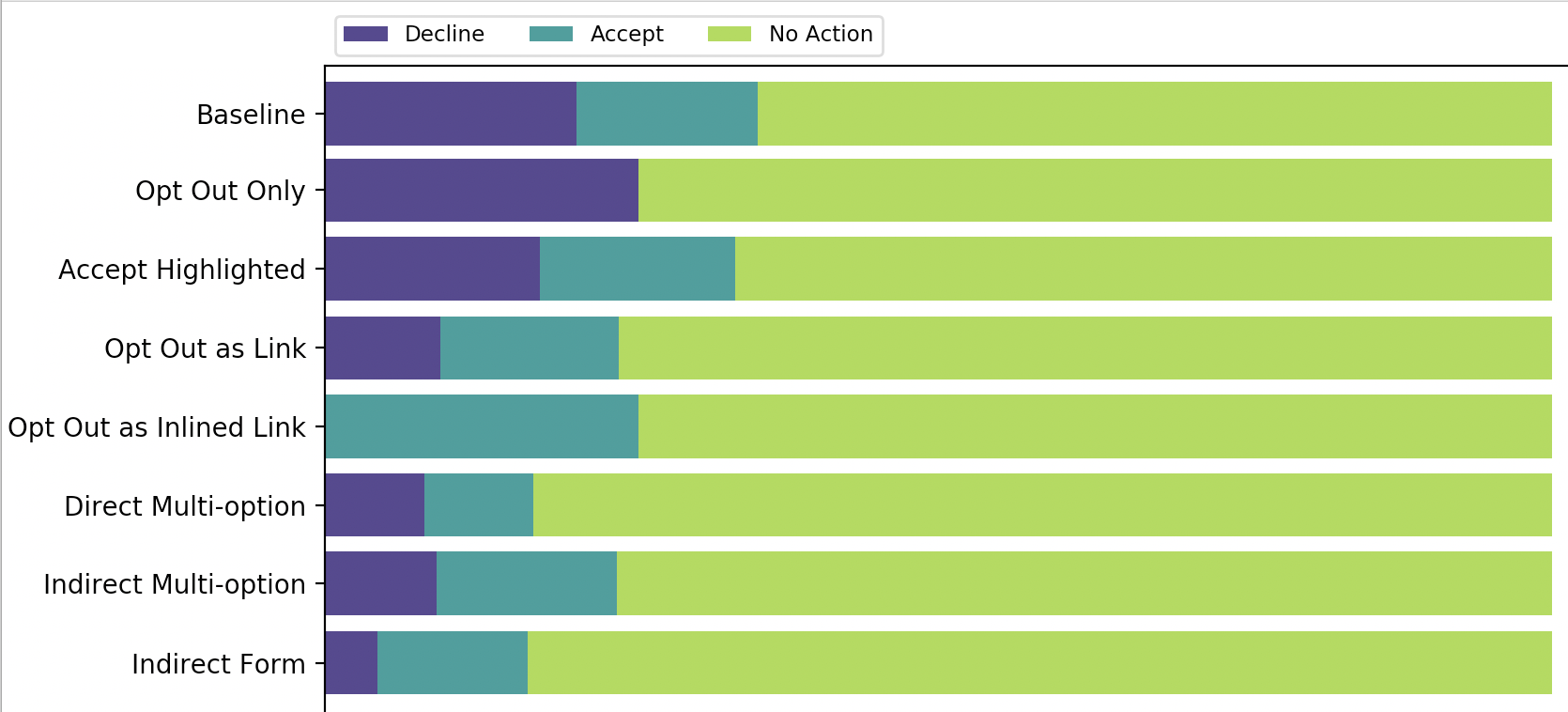}

\caption{The effect of nudging and inconvenience on opt-out interaction rates. }\label{fig:interaction-nudging}
\end{figure}

We also tested three designs incorporating commonly seen inconvenience factors.  First, we implemented a banner requiring the user to separately toggle off three different categories of information sale, instead of a single opt-out button.  We also tested an indirect version of this design, where users were first presented with the baseline banner design, but clicking ``Do Not Sell My Information" led to a second dialogue that presented them with the three-toggle mechanism to complete the process.  Finally, we tested another indirect banner that required users clicking the opt-out button to subsequently fill out a form with their email address.


We found that eliminating the ``Accept'' button (our anti-nudging design) reduced the overall interaction rate but significantly increased the opt-out rate from 20.5\% to 25.6\% ($p < .001$). The nudging design with highlighting did not significantly affect the interaction rate ($p = .302$). However, replacing the opt-out button with an opt-out link in the same location did significantly reduce the interaction rate ($p = .001$) and replacing the opt-out button with an inline ``More Info'' link both significantly reduced the interaction rate and precluded any users from exercising their right to opt-out of sale ($p < .001$). 

We found that all three inconvenience factors significantly reduced interactions with the opt-out mechanism ($p < .001$). If users were asked to individually opt-out of sale of data for different purposes (e.g., personalized ads, analytics, and content personalization), only 8.2\% exercised their right to opt-out of sale for one or more purposes (compared to the baseline opt-out rate of 20.5\%). The condition in which users had to fill out a form in order to exercise their right to opt-out---a relatively common design in the wild---had a particularly low rate of users opting-out at just 4.3\%. These results, summarized in Figure~\ref{fig:interaction-nudging}, are consistent with prior work on the effect of nudging on GDPR consent notices.
\section{Discussion}

Our results provide insight into the impact of various implementation choices for opt-out mechanims under CCPA; this results can therefore offer guidance for websites that want to voluntarily improve the privacy of their users through the adoption of a privacy-enhancing implementation of the right to opt-out of sale.  Our findings confirm that companies can enhance user privacy by adding a pop-up banner on their homepage in addition to a link;  ideally, this banner would located either at the top of their homepage or in another location of high visual importance, but any banner designs is likely to enhance privacy relative to a link-only mechanism.  We also recommend avoiding designs that place additional work on users such as indirect mechanisms and multi-option mechanisms to the maximum extent possible; instead, we recommend utilizing a banner with a single in-banner button that allows users to entirely opt out in one click.  If information needs to be collected from users during opt-out, we recommend minimizing the number and granularity of data collected.  

Our findings can also provide guidance for how to improve the current CCPA and how to write more effective future privacy regulations. First, we recommend that CCPA regulations should expand to regulate opt-out mechanisms beyond the requiring a link.  Information collection from the user should be required to be minimized, and users should never have to input information that the company doesn't already have.  Requiring e-mail for a ``verifiable" request or requiring a full physical mailing address to confirm California residency are examples of such practices.  When no additional user information is needed to implement the opt-out, opt-out should be able to be completed in a single click; navigating to separate pages, requiring users to opt-out for different purposes separately, and displaying vague lengthy disclosures should not be part of the opt-out process.  Users should also not have to opt-out for multiple purposes separately; opt-out should be universally completed in one step.  To be fully effective, regulations should require that opt-out can be completed directly within the site, rather than directing users to a third-party page, telling them to change their browser settings, or requiring them to contact the company to file a request.

Second, our results suggest that regulators need to empower a robust enforcement agency to enforce the ``clear and conspicuous" requirement for Do Not Sell links.  Our observational studies found numerous examples of websites with missing, incorrectly placed, incorrectly worded, or broken Do Not Sell links.  
Moreover, many websites that do have correct links adopt a design that appears to violate the spirit of the ``clear and conspicuous'' requirement. For example, many links are not currently required to be immediately visible without scrolling down on the page, an issue we found present for an overwhelming majority of notices.  Other issues unaddressed by current CCPA regulations include links being hidden in menus, displayed in small fonts, or shown in low-contrast colors. As we have seen in response to revised GDPR guidelines~\cite{machuletz2020multiple}, companies are likely to respond to modified regulations by creating new deceptive designs; effective enforcement of CCPA will therefore depend on robust, flexible, and active litigation to enforce the requirements of CCPA and to develop relevant case law about what constitutes a clear and conspicuous opt-out mechanism.

Finally, our results suggest that privacy regulations that depend on individual opt-out may be inherently flawed. Across all conditions in our user studies, most participants who expressed discomfort with our example site selling their data did not actually utilize the opt-out mechanism on our site. This reinforces prior findings that opt-out mechanisms discourage engagement, a problem that is exacerbated in the online ecosystem due to the large number of individual sites and data brokers that collect personal information about users (for example, there are over 200 data brokers registered in California~\cite{mahoney2020california}, in addition to the potentially hundreds or thousands of websites with whom a user has direct relationships).  To completely opt out from the sale of their information, consumers would have to separately file opt-out requests with every one of these companies.  Given the current state of compliance with CCPA observed in our studies, entirely preventing the sale of one's data remains infeasible.  As such, efforts that go beyond the implementation of CCPA's sale opt-out, such as the adoption of Do Not Sell browser signals or future legislation to limit data sale, will be critical for enhancing user privacy at scale.

\section{Related Work}
Recent privacy regulations, notably CCPA and GDPR, have given rise to questions about how these regulations impact user privacy and how future regulations might further enhance privacy. This line of work takes place within the context of a larger body of work that has explored how aspects of user design affect user engagement in general and interactions with privacy mechanisms in particular.
  
\subsection{Privacy Regulation Compliance in the Wild}

Due to the recent adoption of CCPA, previous work studying the implementation of CCPA has been limited. A Consumer Reports study~\cite{mahoney2020california} asked users to attempt to opt-out of sale on 216 websites from the California Data Broker registrar (all of whom sell data and are subject to CCPA). They found that 11.1\% of data broker websites lacked a CCPA-required homepage Do Not Sell link, and they documented examples of difficult, unclear, and time-intensive opt-out processes. However, they did not perform a comprehensive analysis top websites comprehensively or study how CCPA-compliance has evolved longitudinally; to the best of our knowledge, this work is the first to provide a comprehensive, longitudinal analysis of how websites implement CCPA in the wild. 

Prior work has also considered similar questions about the impact of earlier privacy regulations, notably GDPR. An observational study conducted in May 2018 (immediately after the GDPR went into effect) found that 62.2\% of top European websites displayed cookie notices~\cite{degeling2018we}; however, in 2019 more than 95\% of banners failed to meet GDPR requirements by offering users no or insufficient choices~\cite{utz2019informed,sanchez2019can}, and in 2020 just 11.8\% of banners analyzed met minimal GDPR requirements~\cite{nouwens2020dark}. Observational studies found that most notices were implemented as  bottom bar banners~\cite{utz2019informed}, and that the use of dark patterns in GDPR notices to nudge users towards consent is common~\cite{utz2019informed,nouwens2020dark}. Our work conducts an analogous series of observational studies to understand how CCPA has been implemented in the wild. 

 \subsection{Usability of CCPA Opt-out Mechanisms}

 Cranor et. al. performed a series of three studies examining how different taglines and icons influence user comprehension and recall of Do Not Sell links~\cite{cranor2020design,cranor2020user,cranor2020ccpa,cranor2021informing}. They found that most participants failed to notice Do Not Sell links in an image of a website, that users expect links to opt them out in a single click, and that an overwhelming majority of participants were unaware of CCPA and/or misunderstood what kinds of personal information were included in Do Not Sell---all findings that are consistent with our results; they recommended the adoption of standardized icons and placement, along with enforcement of the existing requirement for standardized language for opt-out links. However, their work did not compare the usability of current links or privacy icons with opt-out banners, and it did not consider the impact design choices after the initial opt-out link on usability. 
 
 The Consumer Reports study~\cite{mahoney2020california}, which asked users to attempt to opt-out of sale on 216 websites from the California Data Broker registrar, also studied the usability of opt-out mechanisms adopted by those sites. In their study, three users were asked to attempt to opt-out of sale of their personal data on each site. They found that  31.4\% of the sites studied displayed their link in such a manner that at least one out of three users was unable to find it, that more than a third of participants spent over five minutes opting out (with a maximum time of over an hour), and 14\% ultimately were unable to successfully complete the process. However,  the limited sample size (3 users per website) and the differences in design choices adopted by each site precluded any statistically significant results about the impact of the observed designs on users' awareness of (and likelihood of invoking) their right to opt-out of sale. 
   
 Earlier work also consistently found that opt-out mechanisms were hard for users to understand and use~\cite{habib2020s,habib2019empirical,leon2012johnny,utz2019informed}. However, those studies were conducted prior to the adoption of CCPA and focused on the usability of opt-out mechanisms under earlier laws, such as the CAN-SPAM Act and GDPR. 

\subsection{Effect of UI on Privacy Notice Usability}

A large body of work has explored the effect of various factors of user interface (UI) design on how users interact with privacy notices and options. 

\paragraph{Banner Position.} A large-scale experiment varying the position of GDPR cookie consent notices found that users were significantly more likely to interact with banners in the lower left corner and less likely to interact with banners at the top of the page~\cite{utz2019informed}, a result that is inconsistent with our finding that users interact more with banners located at the top of the page. However, that study focused on European users (who frequently encounter consent notice banners and may be trained to look for them in particular locations). The different results might also be due to differences in time spent on the page (that study used an ecommerce site where most users stayed for only a few seconds whereas our study used a news site where the average users visited for 47 seconds) or due to difference in banner size and design (the effect of banner position has been extensively studied in the context of advertising banners, with inconsistent results~\cite{cantoni2013banner, lim2000position, doyle1997banner,nielsen2007banner,garcia1991eyes,adam2007eyetracking}). 

\paragraph{Blocking.} Blocking the underlying website has also been shown to increase user interaction with cookie banners by 3.8 times in another experiment~\cite{nouwens2020dark}, a result consistent with prior work on browser phishing warnings~\cite{egelman2008you}. We found that blocking mechanisms were uncommon in the wild, so we did not evaluate its effect on engagement with CCPA opt-out mechanisms. 

\paragraph{Nudging.}
The use of design elements to prevent the user from making privacy-friendly choices is part of a broader literature on web design dark patterns~\cite{council2018deceived}.  Substantial work has gone into developing a taxonomy for Dark Patterns~\cite{conti2010malicious, brignull2019dark, bosch2016tales, gray2018dark}, which exists within a larger literature on nudging~\cite{acquisti2017nudges}. Prior work has found that dark patterns are commonly used by ecommerce sites to encourage users to make more purchases and disclose more information~\cite{mathur2019dark}; our work extends this finding by observing the prevalence of various types of nudging among CCPA opt-out mechanisms. 

Prior work has consistently found that nudging and dark patterns effectively influence user behavior, including causing users to share more information on social networks~\cite{knijnenburg2014increasing}, and that defaults in particular cause users to accept more cookies~\cite{utz2019informed,nouwens2020dark}. In recognition of these effects, GDPR (unlike CCPA) bans specific anti-privacy designs in cookie consent notices, such as pre-selected checkboxes~\cite{nouwens2020dark}. However, some privacy advocates have argued that nudging should be use to nudge users towards privacy-protecting choices~\cite{acquisti2013gone,schaub2015design}. This work extends these CCPA opt-out mechanisms by finding that certain forms of nudging can either increase or decrease the rate of user opt-out.

\paragraph{Inconvenience Factors.}
Studies have also found the effort required by privacy choices can pose a barrier to users.  A study of Data Deletion and Targeted Advertising Opt-Out Choices~\cite{leon2012johnny} found that opt-out implementations frequently required excessive work from users (26.1-37.5 actions on average
).  They found that users had difficulty completing some task, such as writing an email request to opt-out, and that no users were willing to utilize multiple third party opt-out pages in order to opt-out. Among direct mechanisms, users are less likely to interact with fine-grained mechanisms than with binary choices in cookie notices~\cite{utz2019informed}; fine-grained privacy settings for social network increase users' regret about their choices and decrease user satisfaction~\cite{knijnenburg2014increasing, korff2014too}, although fine-grained privacy settings for location can made users feel more comfortable, suggesting the effect of granularity depends on context~\cite{tang2012implications}. This work is the first to study the effect of inconvenience factors (including fine-grained vs. binary mechanisms) on user engagement with CCPA opt-out mechanisms. 
\section{Conclusion}

The California Consumer Privacy Act (CCPA) has been celebrated as ushering in a new era of privacy protections in the United States. 
This paper identifies 
how CCPA's right to opt-out of sale has been implemented among top US websites, and how implementation has evolved over the first six months of enforcement. 
We then describe a pair of user studies that evaluate the effect of these implementation choices  and  find they negatively impact user privacy. 
These results demonstrate the importance of regulations that provide clear guidelines backed by robust enforcement agencies in order to empower users to exercise their privacy rights. 


\bibliographystyle{ACM-Reference-Format}
\bibliography{main}

\newpage
\appendix

\section{Examples of Opt-out Mechanisms in the Wild}

In this Appendix, we provide examples of the different banner designs observed in the manual observational studies conducted in July 2020 and January 2021.

    \begin{figure}[h!]
    \includegraphics[width=\columnwidth]{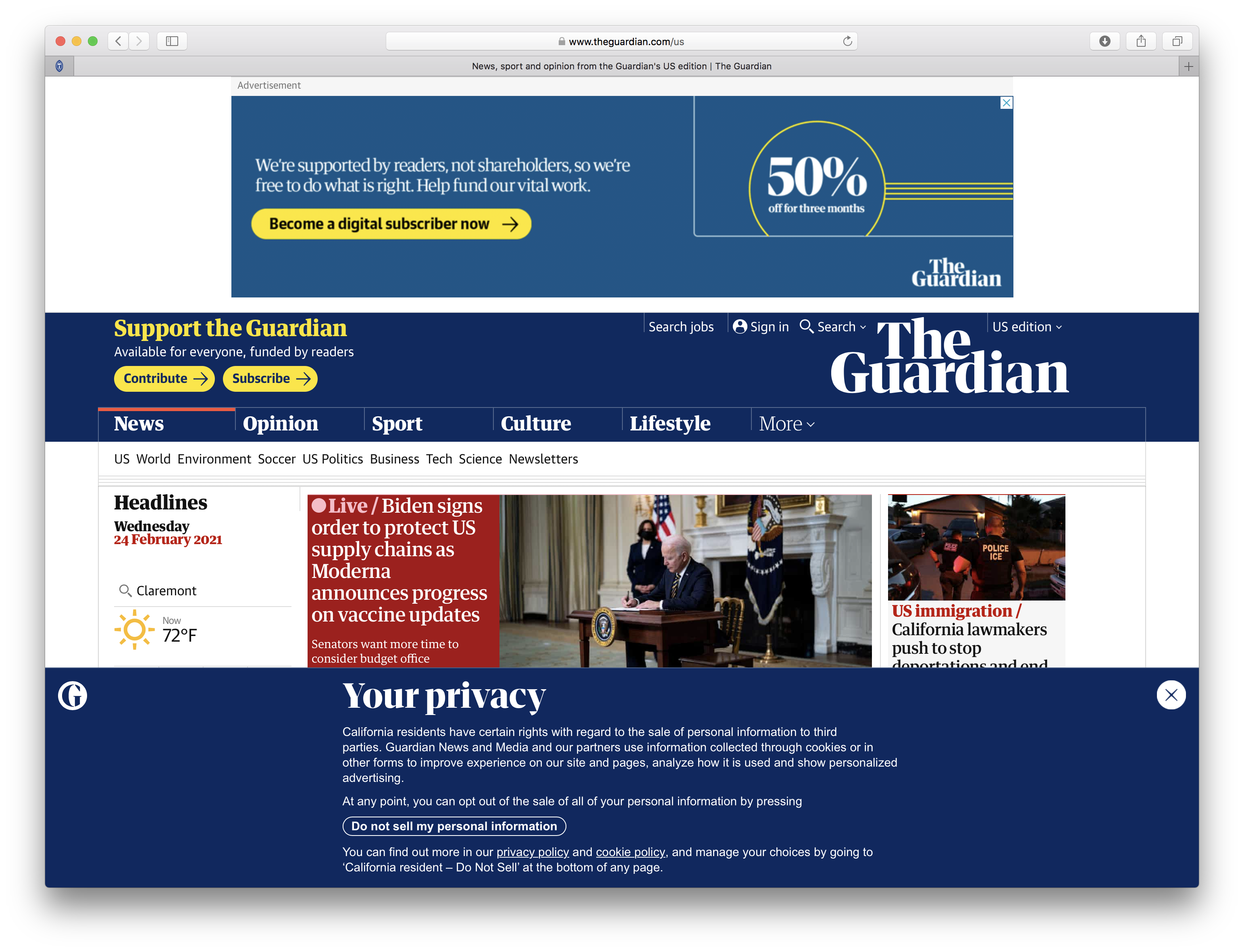}
    \caption{Example website with bottom bar and single in-banner Do Not Sell link.}
    \end{figure}
    \begin{figure}[h!]
    \includegraphics[width=\columnwidth]{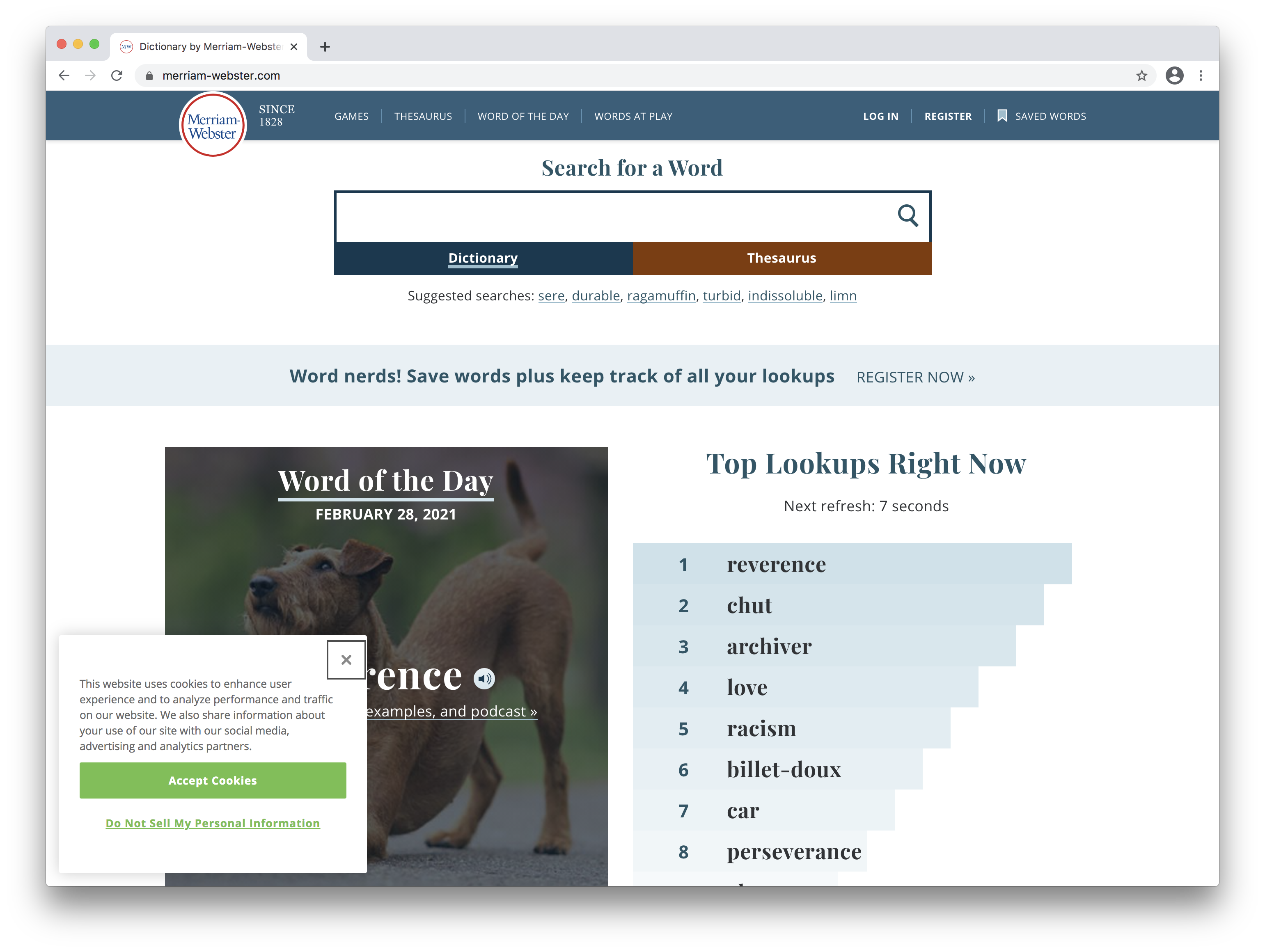}
    \caption{Example website with bottom left banner and two options.}
    \end{figure}
    \begin{figure}[h!]
    \includegraphics[width=\columnwidth,height=2.2in]{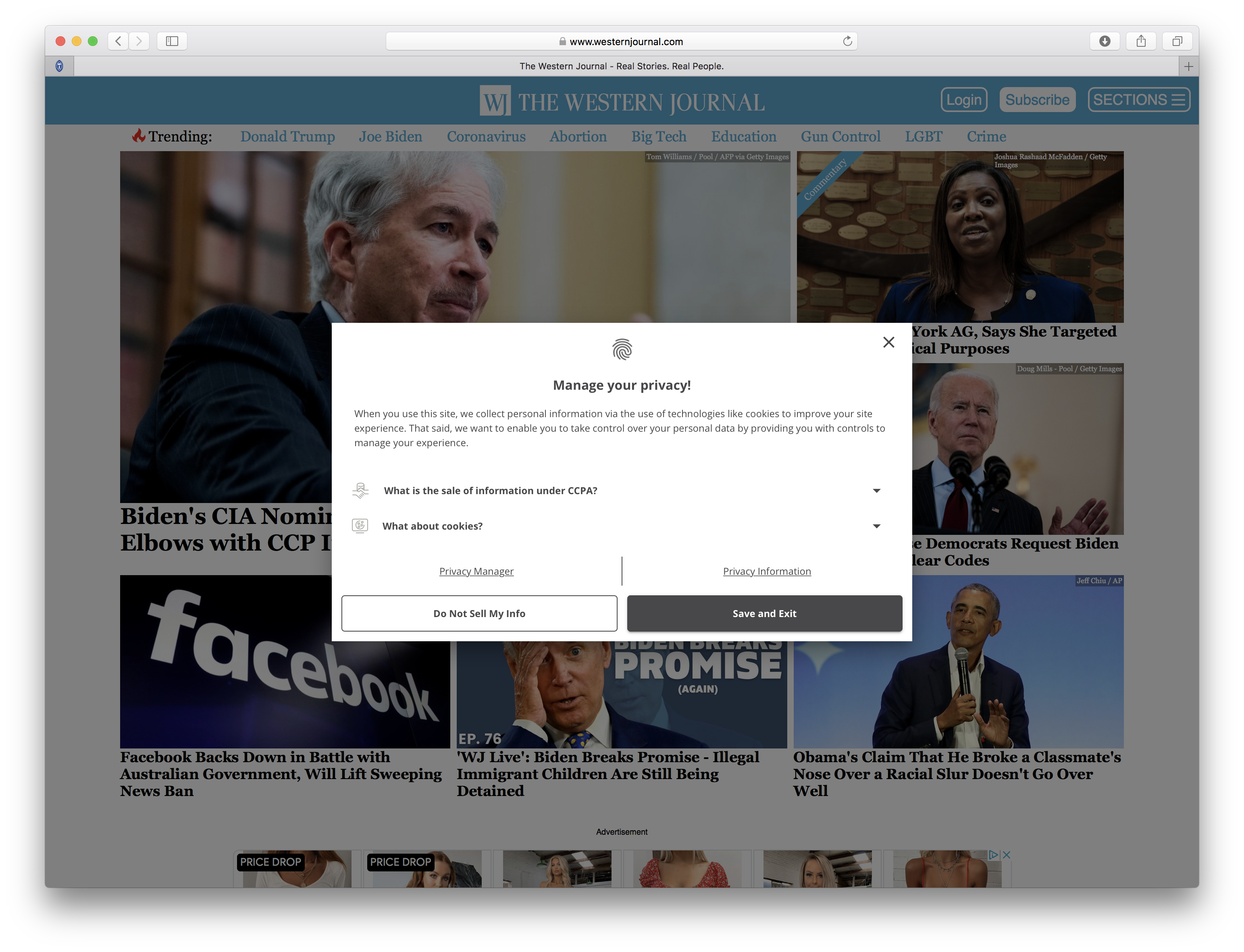}
    \caption{Example website with center banner, two options, and blocking.}
    \end{figure}
    
    \begin{figure}[h!]
    \includegraphics[width=\columnwidth, height=2.2in]{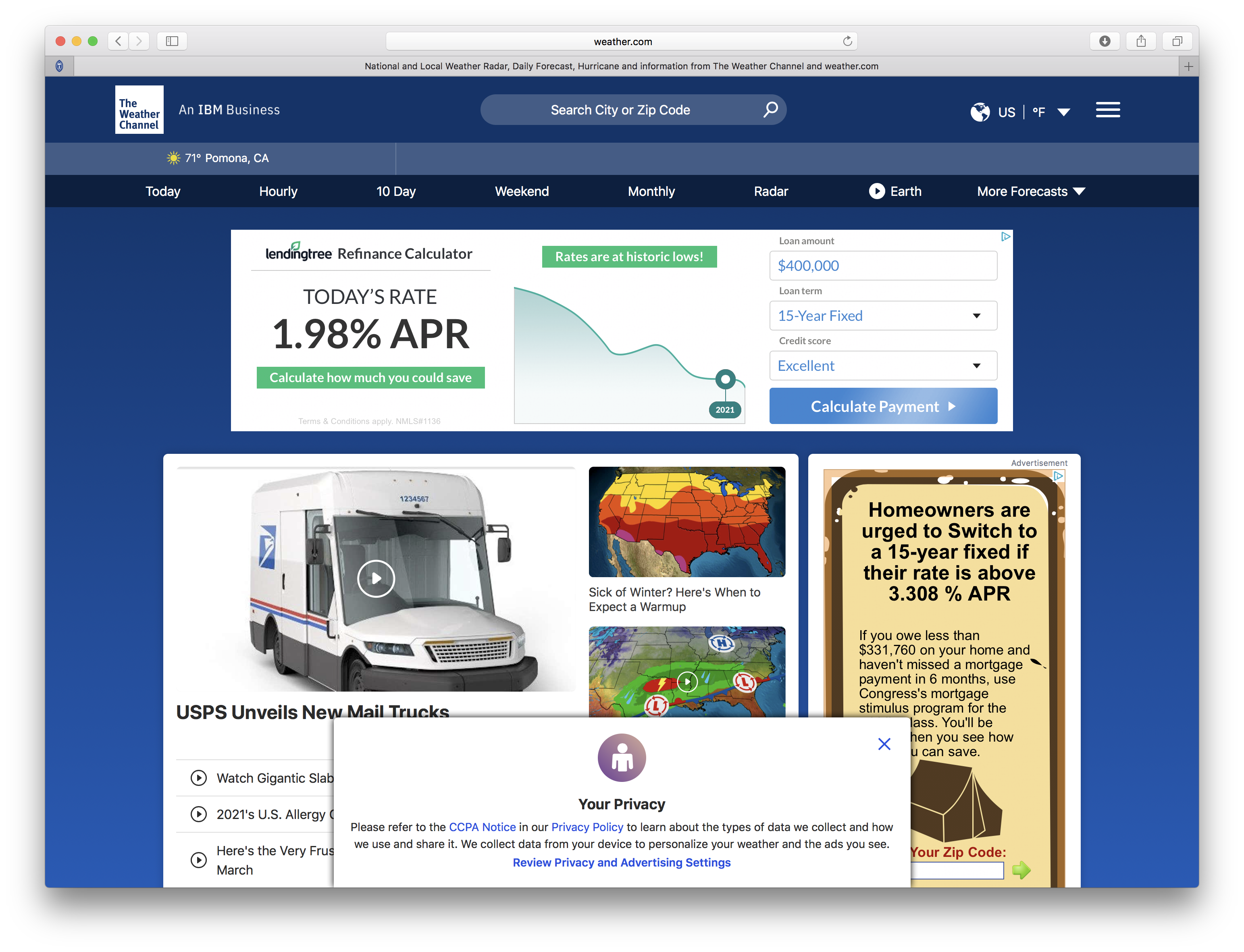}
    \caption{Example website with bottom center banner and opt-out link.}
    \end{figure}
    \begin{figure}[h!]
    \includegraphics[width=\columnwidth,height=2.2in]{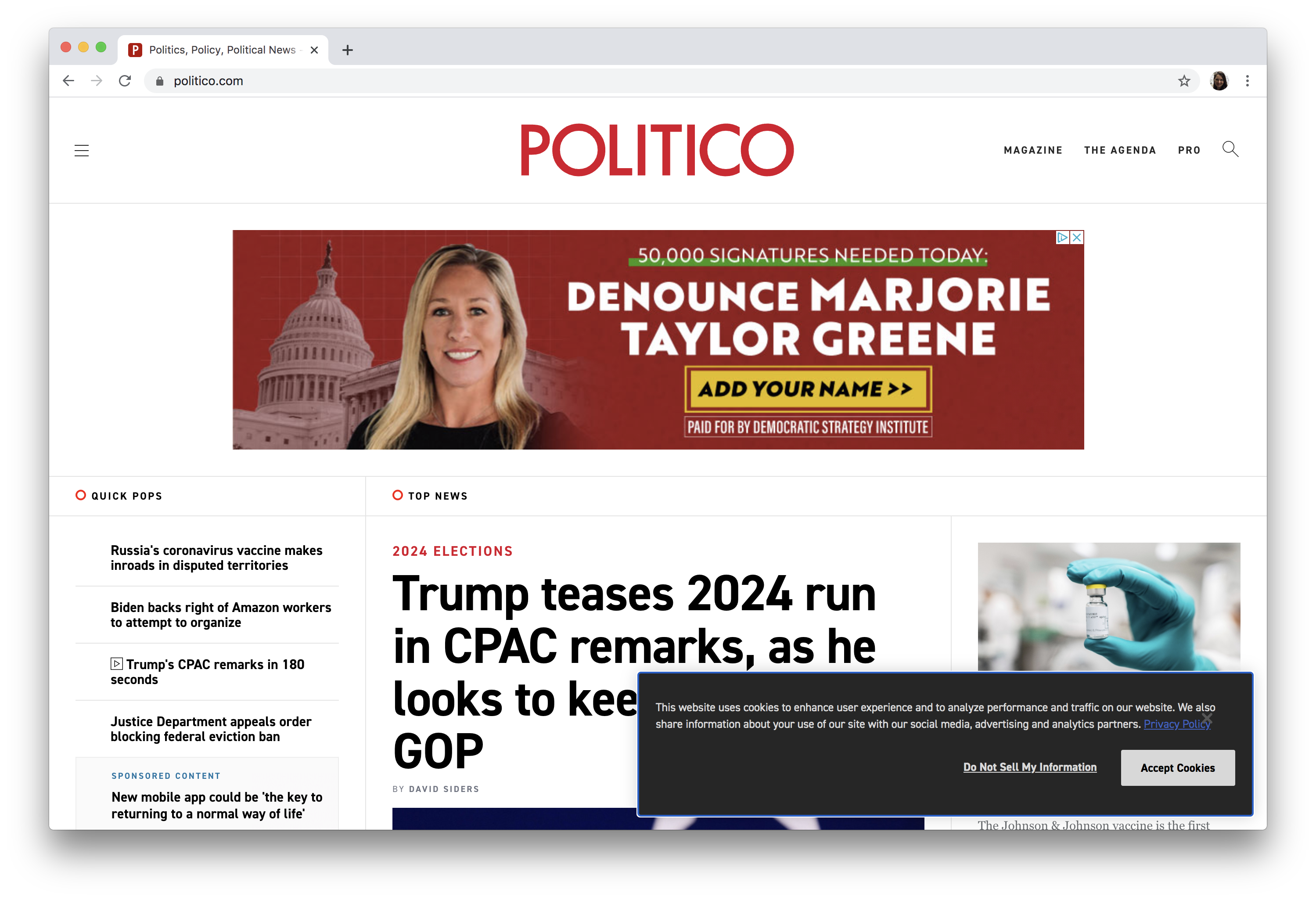}
    \caption{Example website with bottom right banner and two options.}
    \end{figure}

\pagebreak
\section{Follow-up Survey Questions}
In this Appendix, we provide the complete set of questions asked in the follow-up survey provided to study participants recruited through Amazon Mechanical Turk after they had interacted with our example website.

\begin{enumerate}
\item “Did the website you visited track your behavior and sell this information to third parties?” (Yes / No / Unsure)

\item “Did the website you visited give you an option to opt out of the sale of your personal data?”(Yes / No / Unsure)

\item “If this website tracked your behavior and sold this information to third-parties, how comfortable would you be with it?” ( Very Comfortable / Somewhat comfortable / Neutral / Somewhat uncomfortable / Very uncomfortable)

\item “Are you aware that California law requires websites that sell your data to allow you to opt out?” (Yes / No)

\item “How often have you noticed websites you visit giving you an option to opt-out of the sale of your data?” (Never / A few times / Sometimes / Often / Always)

\item “How often do you opt-out of the sale of your data on websites you visit?” (Never Have / Have a few times / Sometimes / Usually / Always)

\item “What is your current age?” (18-24 / 25-34 / 35-44 / 45-59 / 60-74 / 75+)

\item “What is your gender?” (Man / Woman / Non-binary person / Other)

\item “Choose one or more races that you consider yourself to be:” (White / Black or African American / American Indian or Alaska Native / Asian / Pacific Islander or Native Hawaiian / Other)

\item “Do you consider yourself to be Hispanic?” (Yes / No)

\item “How would you describe your political views on social issues?” (Very socially liberal / Somewhat socially liberal / Neither socially liberal nor socially conservative / Somewhat socially conservative / Very socially conservative)

\item “How would you describe your political views on economic issues?” (Very economically liberal / Somewhat economically liberal / Neither economically liberal nor economically conservative / Somewhat economically conservative / Very economically conservative)

\item “In which state do you currently reside?” (50 States / D.C. / Puerto Rico / Not in US)

\end{enumerate}
\end{document}